\PassOptionsToPackage{square,comma,numbers,sort&compress,super}{natbib} 
\documentclass[article,aps, twocolumn,showpacs,amsmath,amssymb,superscriptaddress,nofootinbib,floatfix]{revtex4-2}

\usepackage{color}         
\usepackage[pdftex]{graphicx}      
\usepackage{bm}
\usepackage{natbib}            
\usepackage[colorlinks,plainpages=false,linkcolor=black,urlcolor=black,citecolor=black,pdfpagemode=UseNone,pdfstartview=FitBH]{hyperref}
\usepackage{float}
\usepackage{placeins}
\usepackage{achemso}
\usepackage{tikz}
\usepackage{bbold}
\definecolor{mygreen}{rgb}{0,0.5,0}
\definecolor{mybrown}{RGB}{165,42,42}
\definecolor{mymagenta}{RGB}{255,0,255}
\definecolor{myyellow}{RGB}{255,255,0}
\definecolor{mycyan}{RGB}{0,255,255}

\usepackage{dcolumn}

\usepackage{array,booktabs,tabularx}
\usepackage[caption=false]{subfig}
\usepackage[USenglish]{babel}
\usepackage{braket}
\usepackage{xcolor}
\usepackage{amsfonts}
\usepackage{amssymb}
\usepackage[normalem]{ulem}

\newcommand{\blue}[1]{\textcolor{blue}{#1}}
\begin{document}

\title{Electronic structure and lattice dynamics of 1\textit{T}-VSe$_2$: origin of the 3D-CDW}

\author{Josu Diego}
\email{josu.diego@ehu.eus}
\affiliation{Centro de Física de Materiales (CSIC-UPV/EHU), San Sebastián, Spain}
\affiliation{Fisika Aplikatua Saila, Gipuzkoako Ingeniaritza Eskola, University of the Basque Country (UPV/EHU), San Sebastián, Spain}

\author{D. Subires}
\email{david.subires@dipc.org}
\affiliation{Donostia International Physics Center (DIPC), San Sebastián, Spain}

\author{A. H. Said}
\affiliation{Advanced Photon Source, Argonne National Laboratory, Lemont, IL 60439}

\author{D. A. Chaney}
\affiliation{European Synchrotron Radiation Facility (ESRF), BP 220, F-38043 Grenoble Cedex 9, France}

\author{A. Korshunov}
\affiliation{European Synchrotron Radiation Facility (ESRF), BP 220, F-38043 Grenoble Cedex 9, France}

\author{G. Garbarino}
\affiliation{European Synchrotron Radiation Facility (ESRF), BP 220, F-38043 Grenoble Cedex 9, France}

\author{F. Diekmann}
\affiliation{Ruprecht Haensel Laboratory, Deutsches Elektronen-Synchrotron DESY, 22607 Hamburg, Germany}
\affiliation{Institut f\"{u}r Experimentelle und Angewandte Physik, Christian-Albrechts-Universit\"{a}t zu Kiel, 24098 Kiel, Germany}

\author{S. K. Mahatha}
\affiliation{Ruprecht Haensel Laboratory, Deutsches Elektronen-Synchrotron DESY, 22607 Hamburg, Germany}
\affiliation{UGC-DAE Consortium for Scientific Research, University Campus, Khandwa Road, Indore - 452001, India}

\author{V. Pardo}
\affiliation{Departamento de Física Aplicada, Universidade de Santiago de Compostela, 15782, Santiago de Compostela, Spain}
\affiliation{Instituto de Materiais iMATUS, Universidade de Santiago de Compostela, 15782, Santiago de Compostela, Spain}

\author{J. Strempfer}
\affiliation{Advanced Photon Source, Argonne National Laboratory, Lemont, IL 60439}

\author{Pablo J. Bereciartua Perez}
\affiliation{Deutsches Elektronen-Synchrotron DESY, Notkestr. 85, 22607, Hamburg, Germany}

\author{S. Francoual}
\affiliation{Deutsches Elektronen-Synchrotron DESY, Notkestr. 85, 22607, Hamburg, Germany}

\author{C. Popescu}
\affiliation{ALBA Synchrotron Light Source, 08290 Barcelona, Spain}

\author{M. Tallarida}
\affiliation{ALBA Synchrotron Light Source, 08290 Barcelona, Spain}

\author{J. Dai}
\affiliation{ALBA Synchrotron Light Source, 08290 Barcelona, Spain}

\author{Raffaello Bianco}
\affiliation{Dipartimento di Scienze Fisiche, Informatiche e Matematiche, Universit\`a di Modena e Reggio Emilia, Via Campi 213/a I-41125 Modena, Italy}
\affiliation{Centro S3, Istituto Nanoscienze-CNR, Via Campi 213/a, I-41125 Modena, Italy}
             
\author{Lorenzo Monacelli}
\affiliation{Theory and Simulation of Materials (THEOS), École Polytechnique Fédérale de Lausanne, CH-1015 Lausanne, Switzerland}

\author{Matteo Calandra}
\affiliation{Dipartimento di Fisica, Università di Trento, Via Sommarive 14, 38123 Povo, Italy.}
\affiliation{Sorbonne Universit\'es, CNRS, Institut des Nanosciences de Paris, UMR7588, F-75252, Paris, France}
\affiliation{Graphene Labs, Fondazione Instituto Italiano di Tecnologia, Italy}

\author{A. Bosak}
\affiliation{European Synchrotron Radiation Facility (ESRF), BP 220, F-38043 Grenoble Cedex, France}

\author{Francesco Mauri}
\affiliation{Dipartimento di Fisica, Universit\`a di Roma La Sapienza, Roma, Italy} 
\affiliation{Graphene Labs, Fondazione Instituto Italiano di Tecnologia, Italy}

\author{K. Rossnagel}
\affiliation{Ruprecht Haensel Laboratory, Deutsches Elektronen-Synchrotron DESY, 22607 Hamburg, Germany}
\affiliation{Institut f\"{u}r Experimentelle und Angewandte Physik, Christian-Albrechts-Universit\"{a}t zu Kiel, 24098 Kiel, Germany}

\author{Adolfo O. Fumega}
\affiliation{Department of Applied Physics, Aalto University, 02150 Espoo, Finland}

\author{Ion Errea}
\email{ion.errea@ehu.eus}
\affiliation{Centro de Física de Materiales (CSIC-UPV/EHU), San Sebastián, Spain}
\affiliation{Fisika Aplikatua Saila, Gipuzkoako Ingeniaritza Eskola, University of the Basque Country (UPV/EHU), San Sebastián, Spain}
\affiliation{Donostia International Physics Center (DIPC), San Sebastián, Spain}

\author{S. Blanco-Canosa}
\email{sblanco@dipc.org}
\affiliation{Donostia International Physics Center (DIPC), San Sebastián, Spain}
\affiliation{IKERBASQUE, Basque Foundation for Science, 48013 Bilbao, Spain}

\date{July 2022}

\begin{abstract}
In order to characterize in detail the charge density wave (CDW) transition of 1\textit{T}-VSe$_2$, its electronic structure and lattice dynamics are comprehensively studied by means of x-ray diffraction, angle resolved photoemission (ARPES), diffuse and inelastic x-ray scattering (IXS), and state-of-the-art first principles density functional theory calculations. Resonant elastic x-ray scattering (REXS) does not show any resonant enhancement at either V or Se K-edges, indicating that the CDW peak describes a purely structural modulation of the electronic ordering. ARPES experiments identify (i) a pseudogap at T$>$T$\mathrm{_{CDW}}$, which leads to a depletion of the density of states in the \textit{ML-M'L'} plane at T$<$T$\mathrm{_{CDW}}$, and (ii) anomalies in the electronic dispersion reflecting a sizable impact of phonons on it. A diffuse scattering precursor, characteristic of soft phonons, is observed at room temperature (RT) and leads to the full collapse of the low-energy phonon ($\omega_1$) with propagation vector (0.25 0 -0.3) r.l.u. 
We show that the frequency and linewidth of this mode are anisotropic in momentum space, reflecting the momentum dependence of the electron-phonon interaction (EPI), hence demonstrating that the origin of the CDW is, to a much larger extent, due to the momentum dependence EPI with a small contribution from nesting.
The pressure dependence of the $\omega_1$ soft mode remains nearly constant up to 13 GPa at RT, with only a modest softening before the transition to the high-pressure monoclinic \textit{C2/m} phase. The wide set of experimental data are well captured by our state-of-the art first-principles anharmonic calculations with the inclusion of van der Waals (vdW) corrections in the exchange-correlation functional. The comprehensive description of the electronic and dynamical properties of VSe$_2$ reported here adds important pieces of information to the understanding of the electronic modulations in the family of transition metal dichalcogenides.  

\end{abstract}

\maketitle

\section{Introduction}

The understanding of the origin of the charge density wave (CDW) and its coexistence/competition with other complex phenomena like superconductivity (SC) \cite{Morosan2006,Calandra2011,Leroux2015} or magnetism \cite{acs.jpcc.9b04281, acs.jpcc.0c04913} is one of the main challenges in condensed matter physics. 
Cuprates aside \cite{Ghiri2012,Chang2012,Blanco2012,Blanco2013,LeTacon2014,Miao2018,Miao2019}, layered two-dimensional transition metal dichalcogenides (2D-TMDs) have become the prototypical example for materials with a large variety of ground states that populate their phase diagrams, making them an ideal platform for advanced electronic applications \cite{Akinwande2014}. An intriguing combination of correlated electronic states such as CDWs \cite{wilson}, SC \cite{Clay1971}, excitonic instabilities \cite{Mueller2018}, heavy fermions \cite{Vano2021}, as well as topological \cite{Hsu2017} and spin liquid \cite{Ruan2021} behaviours have been reported both in the bulk and the monolayer limits, despite their relatively simple band structures. 

At a microscopic level, the canonical Fermi surface nesting (FSN) scenario, originally devised to explain the Peierls transition in 1D, is questionable in higher dimensions and the critical role of the EPI has gained more popularity in determining the CDW modulation vector, \textbf{q}$\mathrm{_{CDW}}$ \cite{PhysRevB.77.165135}. Moreover, the reduced electronic screening in 1 and 2-dimensions enhances electronic correlations and drives the CDW transition concomitant with other ordering phenomena. The vivid debate about the CDW origin and its relation with other electronic orders is exemplified in (i) 2\textit{H}-NbSe$_2$, where charge instabilities and SC occur below 33 and 7 K, respectively, and different experimental and theoretical approaches are being discussed \cite{PhysRevB.16.801, varma, Web11Nb, Leroux2015, PhysRevLett.125.106101}; and in (ii) 1\textit{T}-TiSe$_2$, where a long-lasting controversy surrounds the origin and stabilization of the electronic modulation \cite{Web11Ti,Watson2019,Monney2011,Wezel2010} and its possible chiral charge and orbital order \cite{Wezel2011,Castellan2013,Ishioka2010,Ueda2021,Xu2020,Peng2022}. 

The quasi-2D 1\textit{T}-VSe$_2$ is another TMD subject to strong debate and the main focus of this paper. The structure consists of van der Waals (vdW) stacked layers of V atoms covalently bonded to Se within an octahedral coordination, as shown in Fig. \ref{Fig1} (a). A three dimensional (3D) CDW with propagation vector \textbf{q}$\mathrm{_{CDW}}$=(0.25 0 -0.3) r.l.u. emerges below T$\mathrm{_{CDW}}$ = 110 K \cite{Eagle1986}. Soft x-ray ARPES data reported asymmetric dogbone electron pockets centered at the $\mathrm{M}$ and $\mathrm{L}$ points, and assigned the origin of the CDW to a pure electronic mechanism considering that a nesting vector coincides with \textbf{q}$\mathrm{_{CDW}}$ \cite{Stro12}. However, Raman scattering \cite{Pandey2020} and more recent ARPES experiments pointed to a more sophisticated mechanism, involving the coupling of optical phonons to the electronic background, as revealed by kinks in the electronic band dispersion \cite{PhysRevB.104.235137}. The role of the EPI was further corroborated by IXS experiments that unveiled a low-energy damped phonon over a wide range of the Brillouin Zone (BZ) \cite{Diego2021}, in contrast to the sharp dip expected at 2\textbf{k}$\mathrm{_F}$ in the FSN scenario. The latter work also showed that the temperature dependence of the soft mode is critically determined by both anharmonicity and the weak vdW interactions, collapsing at T$\mathrm{_{CDW}}$ following a mean field behaviour with critical exponent $\beta\sim\frac{1}{2}$. This work also demonstrated that the momentum dependence of the EPI is the main driving force of the CDW formation. Furthermore, detailed transport experiments under pressure reported an anomalous enhancement of the CDW ordering temperature up to practically room RT, in stark contrast to the usual pressure induced suppression of the charge modulations observed in other TMDs \cite{Leroux2015}, high T$\mathbf{_c}$ cuprates \cite{Souliou2014}, and the recently discovered kagome metals \cite{Chen2021,Wang2021}. At pressures P$>$13 GPa, a structural transition towards a \textit{C2/m} symmetry phase has been reported, concomitant with the emergence of superconductivity \cite{Sahoo2020}. The complex picture reported in bulk VSe$_2$ extends to the monolayer limit, where distinct competing charge orders with a significant range of transition temperatures have been observed \cite{mon1,mon2}, a behaviour that is explained by the subtle anharmonic condensation of different phonon modes \cite{Fumega2023}. Interestingly, the formation of the CDW in the VSe$_2$ monolayer also competes with ferromagnetism \cite{Fum19}. 
 
In order to shed more light into the origin and interactions determining the CDW in VSe$_2$, here we present a comprehensive experimental and theoretical survey of the properties of 1\textit{T}-VSe$_2$. We combine resonant elastic x-ray scattering, ARPES, diffuse and IXS, and state-of-the-art anharmonic calculations to fully characterize the temperature and pressure dependence of its electronic and lattice dynamical properties. We find that, unlike 1\textit{T}-TiSe$_2$, the static charge modulations do not resonate at either the V or Se K-edge, and the main contributor to the CDW formation is a momentum-dependent anisotropic EPI. 
In addition to the the collapse of the low-energy phonon mode at \textbf{q}$\mathrm{_{CDW}}$ and T$\mathrm{_{CDW}}$, the longitudinal and transverse acoustic branches also undergo weak phonon anomalies throughout the CDW phase transition, highlighting the 3D nature of the charge correlations. Furthermore, the critical phonon does not soften as one heads towards the room temperature, high pressure transition indicating where the transition from \textit{P$\overline{3}$m1} to the \textit{C2/m} symmetry \cite{Guo2021} is of first-order type. We explain how the phonon anomalies of VSe$_2$ are well captured when anharmonic effects are included in the calculations with an appropriate inclusion of vdW forces. 
Taken as a whole, our work demonstrates that the simplicity of the VSe$_2$ structure, with a minimal number of atoms per unit cell, allows a direct comparison between experiments and theory, providing a clear understanding of the origin of the electronic modulations in TMDs. We believe our results can be of inspiration to the understanding of collective ground states in more complex materials, such as high T$_c$ superconducting cuprates \cite{Ghiri2012,Chang2012,Edu2014,Comin2014} and the recently discovered correlated topological kagome metals \cite{Ortiz2020,Zhao2021,Li2022}. 

The paper is organized as follows. In section \blue{II} we describe the experimental and computational details as well as the fundamentals of the Stochastic Self-Consistent Harmonic Approximation (SSCHA) method used to calculate the anharmonic phonons. In Section \blue{III} we present the experimental data and provide comparisons with first-principles calculations. Finally, our conclusions are summarized in Section \blue{IV}.

 \section{EXPERIMENTAL AND COMPUTATIONAL METHODS}
\label{cd}

\subsection{Experimental Methods}
High quality single crystals of VSe$_2$ were synthesized by chemical vapor transport using iodine as transport agent \cite{Diego2021}. The chemical composition was confirmed with energy dispersive analysis (EDX). Resonant elastic x-ray resonant x-ray scattering measurements were taken at the 4-ID-D station at the Argonne Photon Source (APS) and the P09 beamline at DESY. The ARPES experiments, with energy and angular resolution of 20 meV and 0.2$^{\circ}$ respectively, were performed at the LOREA beamline at the ALBA synchrotron. The lattice dynamics of VSe$_2$ were studied by high resolution IXS and thermal diffuse scattering (TDS) at the HERIX spectrometer at the 30-ID beamline of the Advanced Photon Source (APS, $\Delta$E=1.5 meV), Argonne National Laboratory and the ID28 beamline at the European Synchrotron Research Facility (ESRF, $\Delta$E=3 meV), with incoming energies of 23.7 and 17.8 KeV for IXS and TDS, respectively. Orientation matrix refinement and reciprocal space reconstructions from the TDS images were performed using the CrysAlis software package. Final TDS reconstructions were processed using in-house code. The components ($h$ $k$ $l$) of the scattering vector are expressed in reciprocal lattice units (r.l.u.), ($h$ $k$ $l$)= $h \mathbf{a}^*+k \mathbf{b}^*+l \mathbf{c}^*$, where $\mathbf{a}^*$, $\mathbf{b}^*$, and $\mathbf{c}^*$ are the reciprocal lattice vectors. low-energy modes of VSe$_2$ were analyzed by fitting the phonon frequencies to damped harmonic oscillators convoluted with the instrumental resolution.
High-pressure powder XRD measurements up to 20 GPa were carried out at RT at the BL04-MSPD beamline of ALBA synchrotron ($\lambda$=0.4246 \r{A}, beam size 20 $\times$ 20 $\mu$m) \cite{Fauth2013} using a LeToullec membrane-type diamond-anvil cell (DAC), with a mixture of methanol–ethanol (4:1) as the pressure-transmitting medium. The pressure was determined using the equation of state (EOS) of copper (Cu) \cite{Dewaele2004}. The data was analyzed with the GSAS suite \cite{Toby2001}. High pressure IXS experiments were performed at the ID28 beamline of the ESRF with an incident photon energy of 17.8 keV and an helium-pressurized membrane DAC. 

\subsection{Computational Methods}

\subsubsection{Electronic property calculation details}
Band structure calculations were performed within density-functional theory (DFT) using the all-electron, full-potential {\sc wien2k} code \cite{wien} based on the augmented plane-wave plus local orbital (APW+lo) basis set.
The generalized gradient approximation (GGA) in the Perdew-Burke-Ernzerhof (PBE) \cite{PBE} scheme was used for the exchange-correlation functional, with a fully converged \textit{k}-mesh of R$_{mt}$K$_{max}$=7.0 and muffin-tin radii of 2.5 and 2.22 a.u. for V and Se, respectively. The theoretical ARPES spectra were calculated with the Chinook code \cite{Day2019} after \textit{wannierization} of the DFT bands.

\subsubsection{Anharmonic lattice dynamics: the SSCHA method}
\label{dyn}

Vibrational properties of 1\textit{T}-VSe$_2$ are described within the Born-Oppenheimer approximation, so that the ionic vibrations are described by the Born-Oppenheimer energy surface (BOES) $V(\mathbf{R})$. $\mathbf{R}$ specifies the positions of all ions in the crystal, i.e. $R_{a}$ in component free notation. Here, the single index $a = (\alpha, s, \textbf{l})$ indicates Cartesian coordinates $\alpha$, atom index inside the unit cell  $s$, and crystal lattice vector $l$ at the same time. Throughout this manuscript bold letters generically denote vectors and tensors in component-free notation.

The SSCHA \cite{sscha1, PhysRevB.96.014111,sscha3, sscha4, sscha5} is a quantum variational method on the free energy fully accounting for anharmonic effects and valid at any temperature. The variational minimization is carried out with respect to a trial density matrix $\rho_{\mathcal{H}}$ defined by a harmonic Hamiltonian $\mathcal{H}$, so that it contains two groups of parameters: the auxiliary force-constants $\boldsymbol{\Phi}$ and the average ionic positions $\boldsymbol{\mathcal{R}}$, the so-called \textit{centroid} positions. In the minimization procedure, the free energy and its gradient are evaluated through a stochastic technique by calculating forces and energies in supercells, without approximating the ionic potential $V(\mathbf{R})$. The centroid positions at the free energy minimum $\boldsymbol{\mathcal{R}}_{eq}$ are the ionic equilibrium positions fully accounting for quantum, thermal and anharmonic effects. The dynamical matrix based on the final auxiliary force-constants,
\begin{equation} \label{eqn:aux}
    D^{(s)}_{ab}=\frac{1}{\sqrt{M_aM_b}} \Phi_{ab}(\boldsymbol{\mathcal{R}}_{eq}),
\end{equation}
determines the amplitude of the vibrations around the $\boldsymbol{\mathcal{R}}_{eq}$ positions, and is, thus, positive definite by construction. $M_a$ and $M_b$ are the masses of atom $a$ and $b$.

According to Landau’s theory \cite{landau}, second-order structural phase transitions like many CDWs in 2D-TMDs can be identified through the analysis of the temperature evolution of the free energy curvature with respect to the order parameter in a high-symmetry configuration. The appropriate order parameter in displacive second-order phase transitions are the precisely the centroid positions. Within the SSCHA formalism, this free energy Hessian can be calculated analytically as \cite{PhysRevB.96.014111}
\begin{equation}\label{eqn:fcurvature}
	\frac{\partial^{2} F}{\partial \boldsymbol{\mathcal{R}}\partial \boldsymbol{\mathcal{R}}} = 
	\boldsymbol{\Phi}+\overset{(3)}{\boldsymbol{\Phi}} \boldsymbol{\Lambda}(0) [\mathbb{1}-\overset{(4)}{\boldsymbol{\Phi}} \boldsymbol{\Lambda}(0)]^{-1} \overset{(3)}{\boldsymbol{\Phi}} \;,
\end{equation}
where $\overset{(n)}{\boldsymbol{\Phi}}$ are the \textit{n}-th order anharmonic force constants, which are calculated as quantum averages taken with the SSCHA density matrix: $\overset{(n)}{\boldsymbol{\Phi}}= \Big\langle {\frac{\partial^{n} V}{\partial {\textbf{R}}^n }} \Big\rangle_{\rho_{\mathcal{H}}}$. The fourth-order tensor $\boldsymbol{\Lambda}(0)$ is the static limit of $\boldsymbol{\Lambda}(z)$, which in component notation is given by:
\begin{equation}
\begin{split}
	\Lambda&^{abcd}(z)=-\frac{1}{2}\sum_{\mu\nu} F(z,\omega_{\mu},\omega_{\nu}) \\
	& \times \sqrt{\frac{\hbar}{2M_a\omega_{\mu}}}\varepsilon_{\mu}^a
	\sqrt{\frac{\hbar}{2M_b\omega_{\nu}}}\varepsilon_{\nu}^b\sqrt{\frac{\hbar}{2M_c\omega_{\mu}}}\varepsilon_{\mu}^c
	\sqrt{\frac{\hbar}{2M_d\omega_{\nu}}}\varepsilon_{\nu}^d \;,
\end{split}
\end{equation}
where  $\omega^2_{\mu}$ and $\varepsilon_{\mu}^a $ are the eigenvalues and corresponding eigenvectors of the auxiliary SSCHA dynamical matrix $D^{(s)}_{ab}$ in Eq. \eqref{eqn:aux}, and
\begin{equation}
\begin{split}
F(z,\omega_{\mu},\omega_{\nu})=&\frac{2}{\hbar} \left[\frac{(\omega_{\mu}+\omega_{\nu})[1+n_B(\omega_{\mu})+n_B(\omega_{\nu})]}{(\omega_{\mu}+\omega_{\nu})^2-z^2}\right.\\
	&\left.-\frac{(\omega_{\mu}-\omega_{\nu})[n_B(\omega_{\mu})-n_B(\omega_{\nu})]}{(\omega_{\mu}-\omega_{\nu})^2-z^2}\right],
\end{split}
\end{equation}
with $ n_{B}(\omega)$ the bosonic occupation factor. Therefore, negative eigenvalues of the free energy Hessian based dynamical matrix,
\begin{equation}
    {D}_{ab}^{(F)}=\frac{1}{\sqrt{M_aM_b}}	\frac{\partial^{2} F}{\partial \mathcal{R}_a \partial \mathcal{R}_b}\Big|_{\mathcal{R}_{eq}},
    \label{eq:fe_dyn}
\end{equation}
indicate a structural distortion that decreases the free energy. This information can be used to determine CDW transition temperatures, since the CDW will appear when $\mathbf{D}^{(F)}$ develops a negative eigenvalue in the high-symmetry phase upon lowering the temperature.
This technique has been successful in the  characterization of displacive structural phase transitions in a large variety of strongly anharmonic systems, such as superconducting hydrides \cite{sscha3, Errea2020}, thermoelectric compounds \cite{unaiSnSe, unaiSnS}, and similar CDW materials \cite{doi:10.1021/acs.nanolett.9b00504, PhysRevLett.125.106101,Zhou2020Theory,Zhou2020Anharmonicity}. 

The physical phonons measured experimentally with, for instance, inelastic scattering experiments are not the eigenvalues of $\mathbf{D}^{(F)}$ nor $\mathbf{D}^{(s)}$, but the spectral functions associated to the displacement-displacement correlation functions. Within the SSCHA \cite{PhysRevB.96.014111,PhysRevB.103.104305, PhysRevResearch.3.L032017}, for a given point \textbf{q} of the BZ, it is given by \cite{sscha3}
\begin{equation}  \label{eqn:sf}
    \sigma(\textbf{q}, \omega) = -\frac{\omega}{\pi} \text{Im} \left( \text{Tr} [ \omega^2 \mathbb{1} - \textbf{D}^{(s)}(\textbf{q})-\boldsymbol{\Pi}(\textbf{q},\omega+i\delta^{+})]^{-1} \right), 
\end{equation}
where $\boldsymbol{\Pi}(\textbf{q},z)$ is the SSCHA self-energy, defined as
\begin{equation}
\boldsymbol{\Pi}(z)= \textbf{M}^{-\frac{1}{2}}\overset{(3)}{\boldsymbol{\Phi}} \boldsymbol{\Lambda}(z) [\mathbb{1}-\overset{(4)}{\boldsymbol{\Phi}} \boldsymbol{\Lambda}(z)]^{-1} \overset{(3)}{\boldsymbol{\Phi}} \textbf{M}^{-\frac{1}{2}} \;,
\label{eq:piz}
\end{equation}
and $\delta^{+}$ is a small positive number.
Note that in the static $\omega=0$ limit the peaks of the of the $\sigma(\textbf{q}, \omega)$ spectral function coincide with the eigenvalues of $\mathbf{D}^{(F)}$.

As a first approximation in the calculation of the spectral function, the mixing between phonon modes can be neglected by assuming that $\boldsymbol{\Pi}(z)$ is diagonal in the basis of the modes. In this way each  mode (\textbf{q},$\mu$) has a particular contribution to the spectral function:
\begin{equation} \label{nomm}
\begin{split}
    \sigma(\textbf{q}, \omega)=& \sum_{\mu} \frac{1}{2} \left[ \frac{1}{\pi} \frac{-\text{Im}\mathcal{Z}_{\mu}(\textbf{q},\omega)}{[\omega-\text{Re}\mathcal{Z}_{\mu}(\textbf{q},\omega)]^2+[\text{Im}\mathcal{Z}_{\mu}(\textbf{q},\omega)]^2} \right. \\ 
    & \left.+\frac{1}{\pi}\frac{\text{Im}\mathcal{Z}_{\mu}(\textbf{q},\omega)}{[\omega+\text{Re}\mathcal{Z}_{\mu}(\textbf{q},\omega)]^2+[\text{Im}\mathcal{Z}_{\mu}(\textbf{q},\omega)]^2} \right],
\end{split}
\end{equation}
with
\begin{equation} \label{zetamu}
    \mathcal{Z}_{\mu}(\textbf{q},\omega)=\sqrt{\omega_{\mu}^2(\textbf{q})+\Pi_{\mu\mu}(\textbf{q},\omega+i0^{+})},
\end{equation}
where $\Pi_{\mu\mu}(\textbf{q},z)$ is the diagonal part of the self-energy in the mode basis. 

The form of the spectral function in Eq. \eqref{nomm} resembles a superposition of Lorentzian functions, but with frequency-dependent centers and widths, meaning that its actual form  may differ from real Lorentzians. One can take a Lorentzian approximation by fixing the frequency in the $\mathcal{Z}_{\mu}(\textbf{q},\omega)$ function. In the so-called ``one-shot'' approximation the Lorentzian shape is recovered by calculating the position of the peak as 
\begin{equation}
     \overset{(os)}{\Theta_{\mu}}(\textbf{q})=\text{Re}\mathcal{Z}_{\mu}(\textbf{q},\omega_{\mu}(\textbf{q}))
     \label{eq:os_w}
\end{equation}
and the half-width at half-maximum (HWHM) linewidth as
\begin{equation}
    \overset{(os)}{\Gamma_{\mu}}(\textbf{q})=-\text{Im}\mathcal{Z}_{\mu}(\textbf{q},\omega_{\mu}(\textbf{q})).
    \label{eq:os_gamma}
\end{equation}

When the SSCHA self-energy is a small perturbation of the SSCHA free propagator ($\Pi_{\mu \mu} << \omega_{\mu}^2$), a further approximation can be taken by truncating the Taylor expansion of Eq. \eqref{zetamu} at first order:
\begin{subequations} \label{perturbative}
\begin{eqnarray}
 \overset{(pert)}{\Theta_{\mu}}(\textbf{q})=&\frac{1}{2\omega_{\mu}(\textbf{q})} \text{Re}\Pi_{\mu\mu}(\textbf{q},\omega_{\mu}(\textbf{q})),\\
 \overset{(pert)}{\Gamma_{\mu}}(\textbf{q})=&-\frac{1}{2\omega_{\mu}(\textbf{q})} \text{Im}\Pi_{\mu\mu}(\textbf{q},\omega_{\mu}(\textbf{q})).
\end{eqnarray}
\end{subequations}

We have applied the SSCHA theory to the high-temperature $P\bar{3}m1$ phase of VSe$_2$, the normal state phase of 1\textit{T}-VSe$_2$ (Figure \ref{Fig1} (a)), with the experimental lattice parameters at RT: \textit{a} = \textit{b} = 3.35 \AA\ and \textit{c}= 6.09 \AA. 
The Born-Oppenheimer energies and forces required for the SSCHA variational minimization were calculated on 4$\times$4$\times$3 sized supercells (almost commensurate with \textbf{q}$_{\text{CDW}}$) using plane-wave based DFT within the PBE approximation \cite{PBE} of the exchange-correlation functional, making use of the {\sc Quantum ESPRESSO} package~\cite{0953-8984-21-39-395502,0953-8984-29-46-465901}. We also performed calculations by including vdW interactions using Grimme’s semiempirical approach \cite{doi:10.1002/jcc.20495} and within the non-local functional developed by Dion et al. \cite{PhysRevLett.92.246401}. We used an ultrasoft pseudopotential that includes 4$s^2$ 3$d^3$ valence electrons for V and a norm-conserving one with 4$s^2$ 4$p^4$ electrons in the valence for Se. We used a plane-wave energy cutoff of 40 Ry for the wave functions and 450 Ry for the charge density. The BZ integrals were performed in a 3$\times$3$\times$3 k-point grid in the supercell with a Methfessel-Paxton smearing \cite{PhysRevB.40.3616} of 0.01 Ry.

Both static and dynamical SSCHA calculations were performed in the so-called bubble approximation \cite{PhysRevB.96.014111}, setting $\overset{(4)}{\boldsymbol{\Phi}}=0$ in Eqs. \eqref{eqn:fcurvature} and \eqref{eq:piz}.  
The dynamical spectral function was calculated considering a broadening of $\delta^+ = 0.5$ cm$^{-1}$ and a  48$\times$48$\times$32 grid to determine the SSCHA self-energy. The phonon frequencies and third-order force constants at these points were obtained by Fourier interpolation. The spectral function was calculated within the ``no mode-mixing'' approximation, i.e. directly from Eq. \eqref{eqn:sf}, as well as in the ``one-shot'' approximation (Eqs. \eqref{eq:os_w}) and \eqref{eq:os_gamma}, and in the perturbative case (Eq. \eqref{perturbative}).

\subsubsection{Harmonic phonons and the EPI}

Harmonic phonon frequencies and electron-phonon matrix elements were calculated within density functional perturbation theory (DFPT) \cite{RevModPhys.73.515} as implemented in {\sc Quantum ESPRESSO}~\cite{0953-8984-21-39-395502,0953-8984-29-46-465901}. DFPT calculations were performed with the same parameters as in the SSCHA but with a 24$\times$24$\times$16 grid in the unit cell for the BZ integrals and a smearing of 0.005 Ry. The electron-phonon linewidth is given by 

\begin{equation}
    \gamma_{\mu}(\textbf{q})  = \frac{2\pi \omega_{\mu}(\textbf{q})}{N}\sum_{nn'}\sum_{\textbf{k}}^{1BZ}|g^{\mu}_{n'\textbf{k}+\textbf{q},n\textbf{k}}|^2  \delta(\epsilon_{n'\textbf{k}+\textbf{q}})\delta(\epsilon_{n\textbf{k}}),
\label{eq:elph}
\end{equation}
where  $\epsilon_{n\textbf{k}}$ is the energy of band $n$ with wave number $\textbf{k}$ measured from the Fermi level,  $N$ the number of $\textbf{k}$ points in the sum over the first BZ, and $g^{\mu}_{n'\textbf{k}+\textbf{q},n\textbf{k}}$ the electron-phonon matrix elements. In Eq. \eqref{eq:elph} we use a  48$\times$48$\times$32 $\mathbf{k}$-point grid and a Gaussian broadening of 0.003 Ry for the Dirac deltas. 

\section{Results}
\subsection{Characterization of the CDW: transport and resonant elastic x-ray scattering}

\begin{figure} 
\begin{center}
\includegraphics[width=0.8\columnwidth,draft=false]{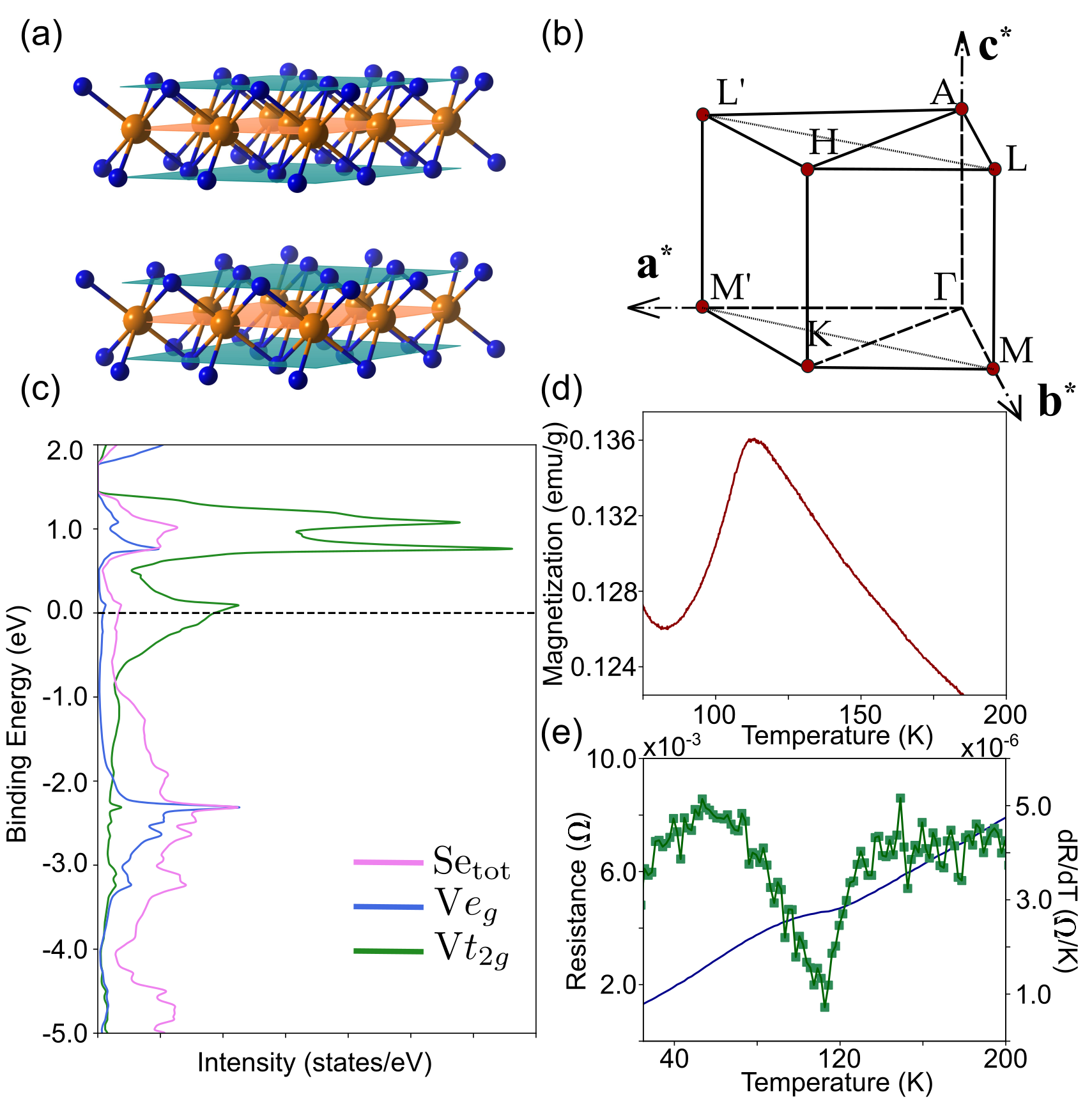}
\caption{(a) 2D structure of VSe$_2$, where V and Se atoms are represented by orange and blue spheres, respectively. (b) BZ of the space group \textit{P$\overline{3}$m1} (164) and the main symmetry directions. (c) Density of states calculated using DFT showing the Se and V character close to the Fermi level. (e) Magnetization and (d) resistance of the VSe$_2$ single crystals, highlighting the CDW transition at 110 K.}
\label{Fig1}
\end{center}
\end{figure}

Figures \ref{Fig1} (a) and (b) show the chemical structure of 1{\it T}-VSe$_2$ and the high symmetry directions of the BZ. The unit cell of the 1-\textit{T} polymorph consists of a 2D stacking of VSe$_2$ monolayers bonded together by vdW interactions \cite{Manzeli2017}. The V ion is octahedrally coordinated by Se atoms within the \textit{P$\overline{3}$m1} space group with lattice parameters \textit{a}= \textit{b}= 3.35 \AA\ and \textit{c}= 6.09 \AA. Transport experiments reveal a CDW transition at 110 K, as shown in Fig. \ref{Fig1} (d-e), in very good agreement with previous reports \cite{Sahoo2020,Tsu82}. The small amplitude of the hump in the resistivity and magnetic susceptibility curves illustrates the high quality of our crystal and demonstrates that only a small fraction of carriers are lost at the CDW transition. Indeed, DFT calculations show that the charge carriers at the Fermi level (Fig. \ref{Fig1} (c)) come mostly from the V bands, with a predominant contribution from the trigonally-split t$_{2g}$ orbitals. The Se p bands dominates from below 1 eV the Fermi level (strongly mixed with the V e$_g$ bands).

\begin{figure} 
\begin{center}
\includegraphics[width=\columnwidth,draft=false]{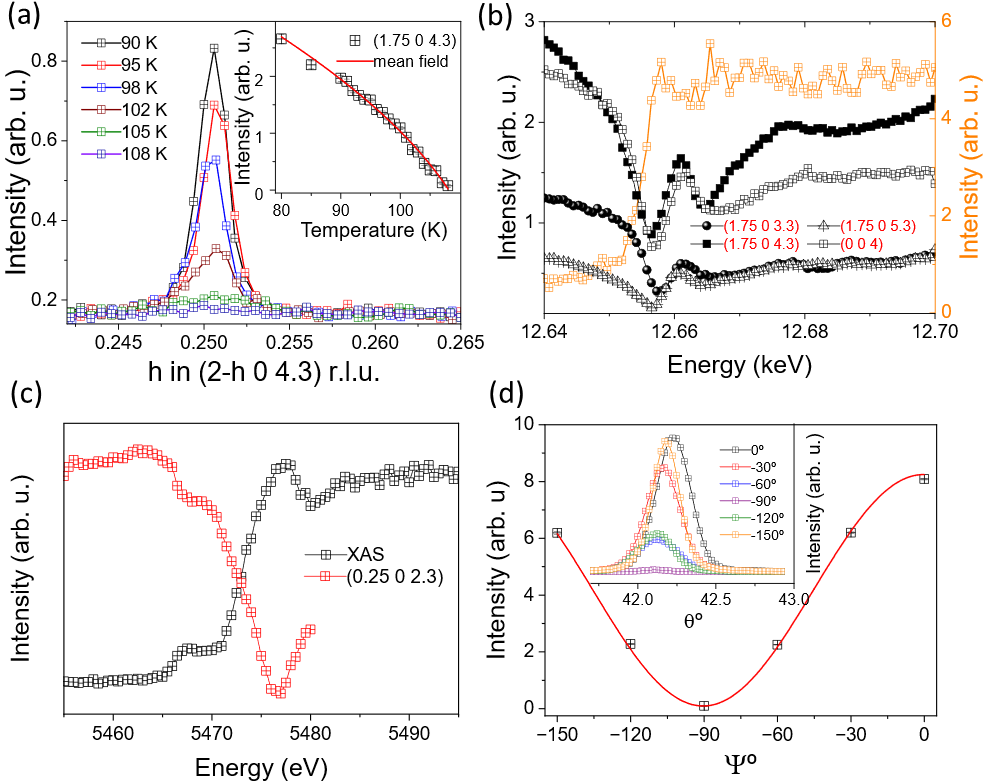}
\caption{(a) Temperature dependence of the CDW peak at the Se \textit{K}-edge (12.66 keV). Inset, fitting of the temperature dependence of the intensity to a power law with $\alpha$=0.43 $\pm$0.05. (b) Energy dependence of several CDW reflections and the (0 0 4) Bragg peak at 60 K. The orange spectrum follows the x-ray absorption at \textbf{q}$\neq$\textbf{q}$\mathrm{_{CDW}}$. (c) Energy scan at the CDW peak (0.25 0 2.3) r.l.u. (red) and absorption spectrum (black) at the V \textit{K}-edge. (d) Polarization dependence of the (0.25 0 2.3) r.l.u peak. Inset, raw spectra.}
\label{Fig2}
\end{center}
\end{figure}

We initiate our experimental survey studying the charge reflections by means of resonant hard x-ray diffraction at the Se and V \textit{K}-edges. Figure \ref{Fig2} (a) displays the momentum-temperature dependence of the raw scans (Se \textit{K}-edge, 12.66 keV) at the (1.75 0 4.3) r.l.u. CDW peak. The integrated intensity follows a mean field power law with $\alpha$=0.43$\pm$0.05, characteristic of a second order phase transition. On the other hand, the energy dependence of the CDW reflections does not show any resonant enhancement of the CDW signal. The charge reflections are dominated by Thompson scattering, as demonstrated by a wiggled energy dependence that follows the (0 0 4) Bragg peak (Fig. \ref{Fig2} (b)). Hence, the resonant scattering described by a full occupation of the Se $\mathrm{p}$ states associated with a purely structural distortion at the CDW transition. Focusing on the V \textit{K}-edge (Fig. \ref{Fig2} (c)), the energy scan displays the main absorption edge at 5.475 keV, which corresponds to the dipole allowed transition from the 1$\mathrm{s}$ to the empty 4$\mathrm{p}$ states of V. A pre-edge absorption is visible 10 eV below the main edge, ascribed to the dipole forbidden 1$\mathrm{s}\rightarrow3\mathrm{d}$ transition, which becomes allowed due to a distorted octahedral symmetry that switches on the mixing between 3$\mathrm{d}$ orbitals of V and 4$\mathrm{p}$ orbitals of Se \cite{Wong1984}. Similar to the Se \textit{K}-edge, the energy scan across the CDW does not show any resonant enhancement at any azimuthal angle pointing to a pure charge scattering process. Fitting the polarization dependence of the scattered radiation (Fig. \ref{Fig2}(d) and inset) with \cite{Detlefs2012},
\begin{equation}
    I\propto S+P_1\cos\left(2\Psi\right)+P_2\sin\left(2\Psi\right)
\end{equation}
returns the values of the Poincaré-Stokes parameters, \textit{P}$_1$ and \textit{P}$_2$ of 0.992 and -0.013, respectively (a value of \textit{P}$_1$ $\sim$ 1 indicates no change in the rotation of the polarized scattered light). This reaffirms the purely structural origin of the CDW reflections. Furthermore, an exhaustive search for signatures of orbital ordering and forbidden reflections by symmetry gave a negative result at both edges. Hence, the CDW of 1\textit{T}-VSe$_2$ seems to be more conventional in comparison with the possible chiral/orbital order reported in the 1\textit{T}-TiSe$_2$ \cite{Ueda2021,Peng2022}.

\subsection{Electronic structure: ARPES}

\begin{figure} 
\begin{center}
\includegraphics[width=0.8\columnwidth,draft=false]{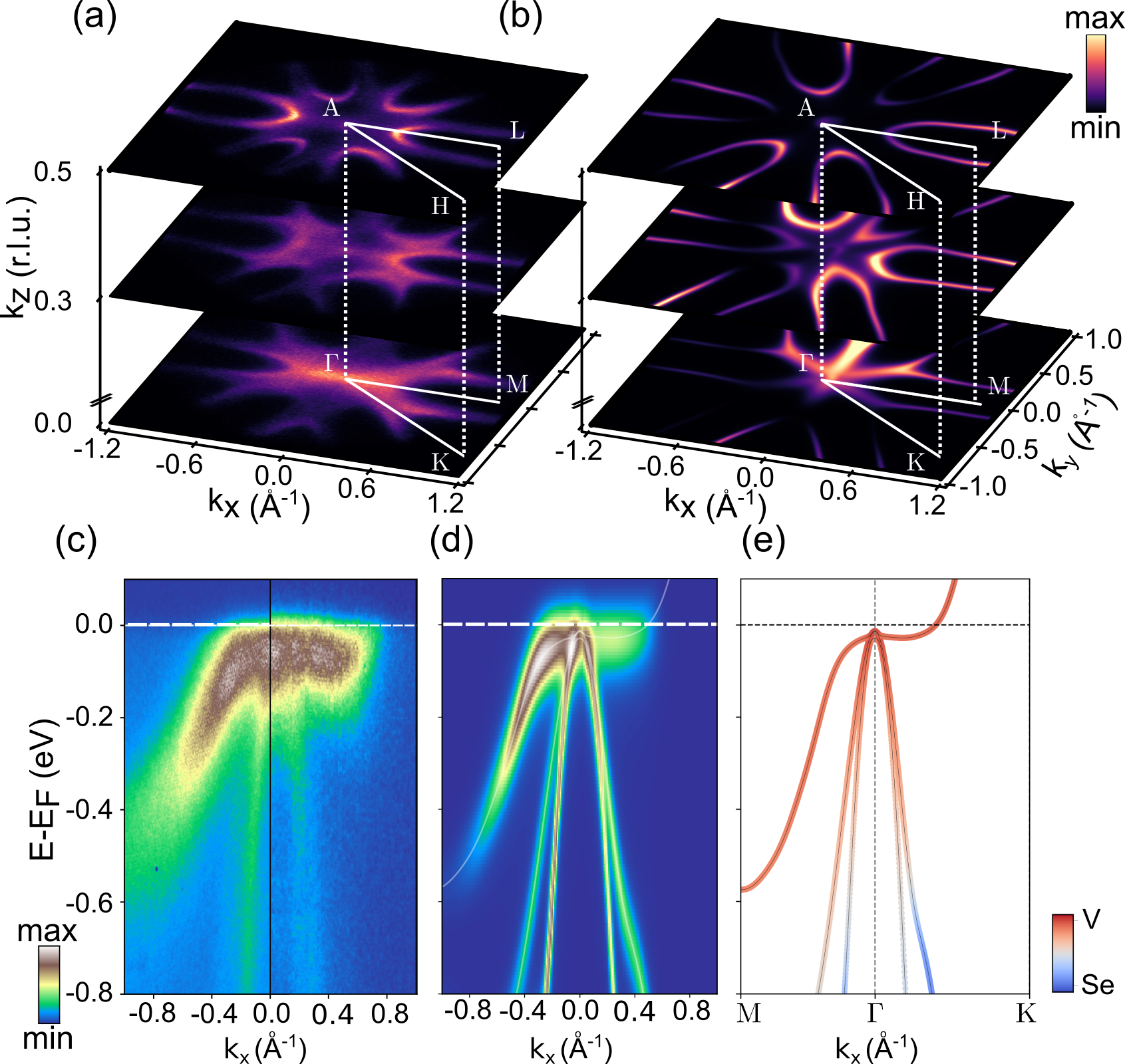}
\caption{(a) Experimental Fermi surface at the $\mathrm{k}_z$= 0, 0.3 and 0.5 r.l.u. planes ($h\nu = 72,\,62,\,57\,\mathrm{eV}$, respectively) in the low temperature phase. (b) ARPES simulated spectra of the Fermi surface at the same planes. (c) Experimental energy-momentum dispersion along the direction $\mathrm{M}-\Gamma-\mathrm{K}$ (\textit{h}$\nu$=57 eV). (d) Simulated energy-momentum dispersion through the direction $\mathrm{M}-\Gamma-\mathrm{K}$. (e) Atomic orbital contribution of vanadium and selenium to the bands in the $\mathrm{M}-\Gamma-\mathrm{K}$ direction. 
}
\label{Fig3}
\end{center}
\end{figure}

We will now detail our study of the electronic structure by means of ARPES. Figure \ref{Fig3} (a) and (b) display an overall nice agreement between the experimental and calculated Fermi surfaces in the $\mathrm{k}_z$= 0, 0.3 and 0.5 r.l.u. planes. At 10 K, the Fermi surface consists of an intense feature at $\Gamma$ and large ellipsoids centered at each $\mathrm{M}$ point that extend over the whole BZ.
In Fig.~\ref{Fig3} (c-d), we plot the measured and calculated band dispersion along the high-symmetry line $\mathrm{M}-\Gamma-\mathrm{K}$.
Fig. \ref{Fig3} (c) shows two nearly degenerate hole-like bands that merge at the Fermi level at the $\Gamma$ point and separate at higher binding energy. The inner of these bands is a largely covalent mixture of Se $\mathrm{p}$ states together with V $e_g$ bands (forming a $\sigma$-bond), while the outer degenerate band comes mostly from the V a$_{1g}$ orbital. This is the only (mostly) occupied t$_{2g}$ orbital in VSe$_2$, with lobes pointing along the trigonal axis of the Se octahedra (off the vdW hexagonal plane of the structure). This band disperses upwards along the high-symmetry line $\Gamma-\mathrm{K}$. The dominant V $\mathrm{d}$ character just below the Fermi level can be seen in Fig.~\ref{Fig3} (e). 
At $\mathrm{A}$, the band character comes mostly from V e$_{g^{\pi}}$ states, the doublet formed by t$_{2g}$ states in a trigonal environment , which is dispersionless along the $\Gamma$-A direction. 
Overall, the measured electronic band structure is fully reproduced by the DFT calculations and is consistent with the previous reports in the literature \cite{Stro12,Yilmaz2022}.

\begin{figure*}
\begin{center}
\includegraphics[width=1.7\columnwidth,draft=false]{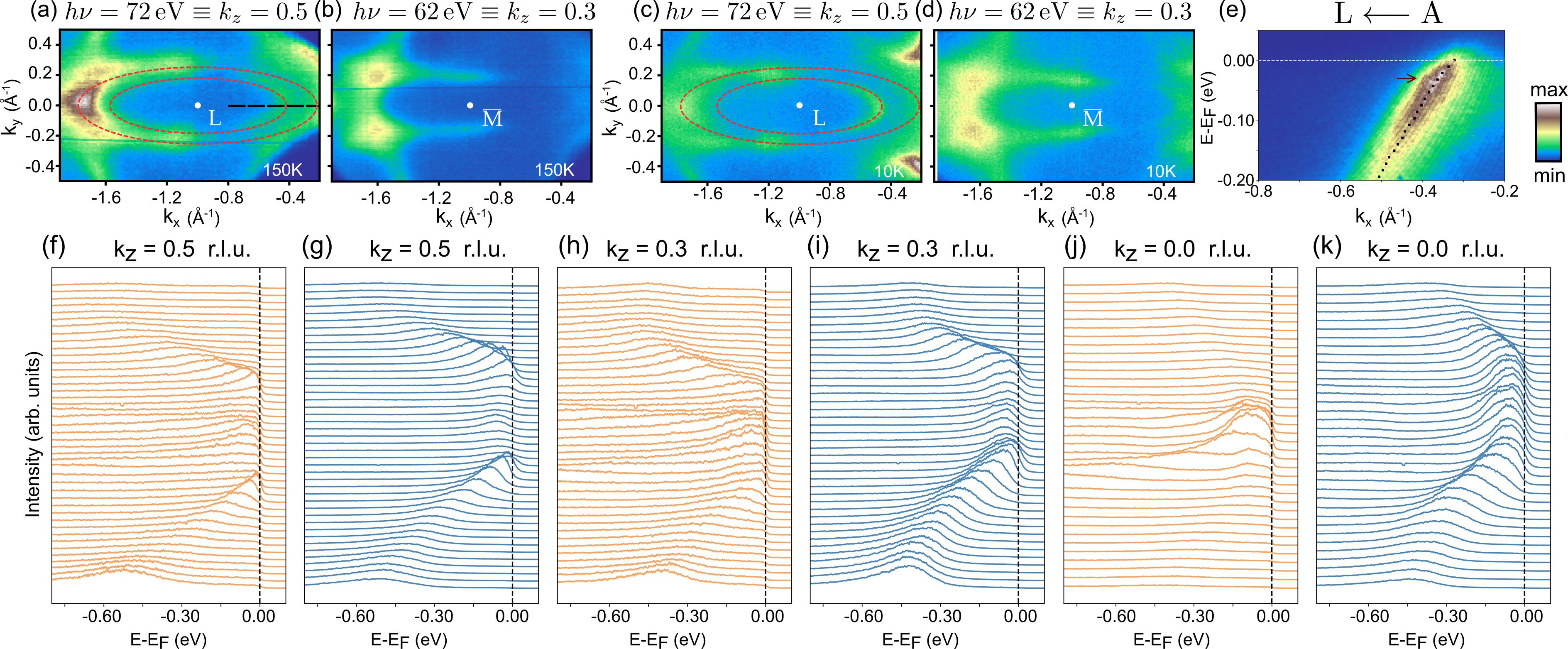}
\caption{ (a), (c) high ((b), (d) low) temperature measured Fermi surface $\pm50\,\mathrm{meV}$ centered on an ellipsoid measured with $h\nu=72,\mathrm{eV}$ ($h\nu=62\,\mathrm{eV}$). The red dashed with ellipses are a guide to the eye. The black dashed line in (a) represent the momentum-dependent energy dispersion.(e) Band dispersion in the direction $\mathrm{L}-\mathrm{A}$ at $T=150\,\mathrm{K}$. The maxima of intensity of the band for each binding energy is marked by the black points and the kink feature is highlighted with an arrow. (f), (h), (j) Energy Distribution Curves (EDCs) at $72,\,62,\,57\,\mathrm{eV}$ at $T=10\,\mathrm{K}$ ((g),(i),(k) at $T=150\,\mathrm{K}$.}
\label{Fig4}
\end{center}
\end{figure*}

On the other hand, the experimental Fermi surface centered at $\mathrm{\overline{M}}$ for $\mathrm{k}_z$=0.3 and 0.5 r.l.u. shows a suppression of spectral weight at the Fermi level around $|\mathrm{k}_x|\sim 0.55$ \r{A}$^{-1}$ ($\textit{h}\sim$ 0.25 r.l.u.) both at low (10 K) and high (150 K) temperature. To better visualize this loss of states, in Fig.~\ref{Fig4} (f-k) we compare the energy distribution curves (EDCs) that characterize the ellipsoidal Fermi surface. Although less states are removed at $\mathrm{k}_z=0$, the loss of intensity happens for all values of $\mathrm{k}_z$, in agreement with the presence of a pseudogap phase above the CDW transition \cite{Ter03}. Further, our ARPES data at low and high temperature resolve 2 ellipsoidal bands around the $\mathrm{\overline{M}}$ point ( Fig.~\ref{Fig4} (a,c)), more clearly visible in the 2nd BZ, which are not captured by the DFT calculations, but are in agreement with recent reports \cite{Yilmaz2022}. Our results also reveal a non-linear band dispersion close to $\mathrm{E_F}$ and a small ``kink'' in the L$\rightarrow$A direction (Fig.~\ref{Fig4} (e)), corroborating the presence of strong EPI, as recently reported by Raman and high resolution ARPES \cite{PhysRevB.104.235137}.

\begin{figure} 
\begin{center}
\includegraphics[width=\columnwidth,draft=false]{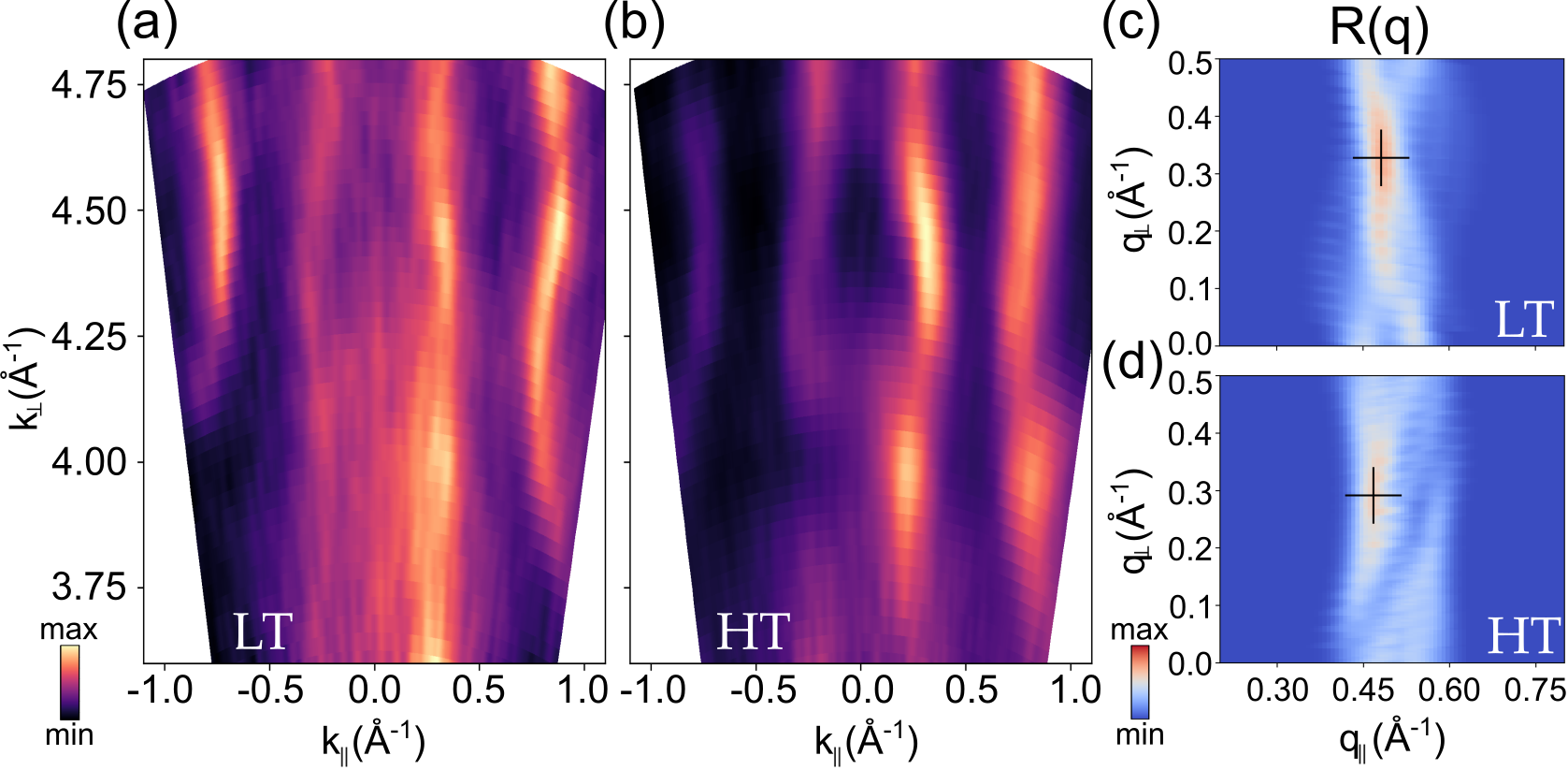}
\caption{(a-b) Fermi surface in the plane $\left(k_\parallel,\,k_\perp\right)$ at $T\simeq10,\,150\,\mathrm{K}$ respectively. The energy scans are plotted at the Fermi Level integrated in a range of $\pm100\,\mathrm{meV}$. (c), (d) Autocorrelation $R\left(q_\parallel,\,q_\perp\right)$ for the $k_z$ dispersion at low and high temperature, respectively. The autocorrelation is maximum at $q_\parallel=0.48\,\mathrm{\AA^{-1}}$ and $q_\perp=0.33\,\mathrm{\AA^{-1}}$ for low temperature and $q_\parallel=0.47\,\mathrm{\AA^{-1}}$ and $q_\perp=0.29\,\mathrm{\AA^{-1}}$ for high temperature. These maximums are indicated with the crosses.
}
\label{Fig5}
\end{center}
\end{figure}

Having confirmed and discussed the existence of a CDW pseudogap, we will now evaluate the presence of Fermi surface nesting vectors. In Fig.~\ref{Fig5}(a-b), we approximate the zero-energy Lindhard charge susceptibility, $\chi_q$, by computing the autocorrelation map of the Fermi surface measured in the $MLL'M'$ plane (see fig.~\ref{Fig1}(b)) \cite{Stro12}. We define the autocorrelation function as $R\left(\bm{q}\right)=\int_\Omega I_F\left(\bm{k}\right)I_F\left(\bm{k+q}\right)d^2\bm{k}$, where $I_F\left(\bm{k}\right)$ is the photoemission intensity of the Fermi surface at each $\bm{k}$. This autocorrelation function has been used extensively to search for a divergent behaviour in $\chi_q$ \cite{Hasimoto2011}. Although our measurements were carried out below $h\nu=100\,\mathrm{eV}$ and the intrinsic k$_z$-broadening is not small enough to treat the final-states as free-electron-like, we can still gain some valuable information using this approach. In Fig.~\ref{Fig5} (c-d), we plot $R\left(\bm{q}\right)$ at low and high temperature, respectively. At $150\,\mathrm{K}$, the autocorrelation map develops a broad streak of intensity with its maximum at $q_\parallel=0.23\,\mathrm{r.l.u.}$ and $q_\perp=0.29\,\mathrm{r.l.u.}$ At low temperature, this maximum appears slightly shifted, ($q_\perp=0.33\,\mathrm{r.l.u.}$), but still very close to the \textbf{q}$\mathrm{_{CDW}}$ observed experimentally in the diffraction experiments, hence, demonstrating the enhanced charge correlations at \textbf{q}$\mathrm{_{CDW}}$. Indeed, the charge correlations appear throughout the whole k$_z$ direction, not exclusively at \textbf{q}$\mathrm{_{CDW}}$. It should be stressed that this does not support the Fermi surface nesting scenario, but correlates with the softening of the low-energy phonon branches, as we will detail later.   

\subsection{Lattice dynamical properties of the high-symmetry phase}

Having studied the structural and electronic properties, we now move to the lattice dynamical properties of  of 1\textit{T}-VSe$_2$. 

\subsubsection{Anharmonic DFT spectral functions}

\begin{figure*}[t]
\begin{center}
	\includegraphics[width=1.62\columnwidth]{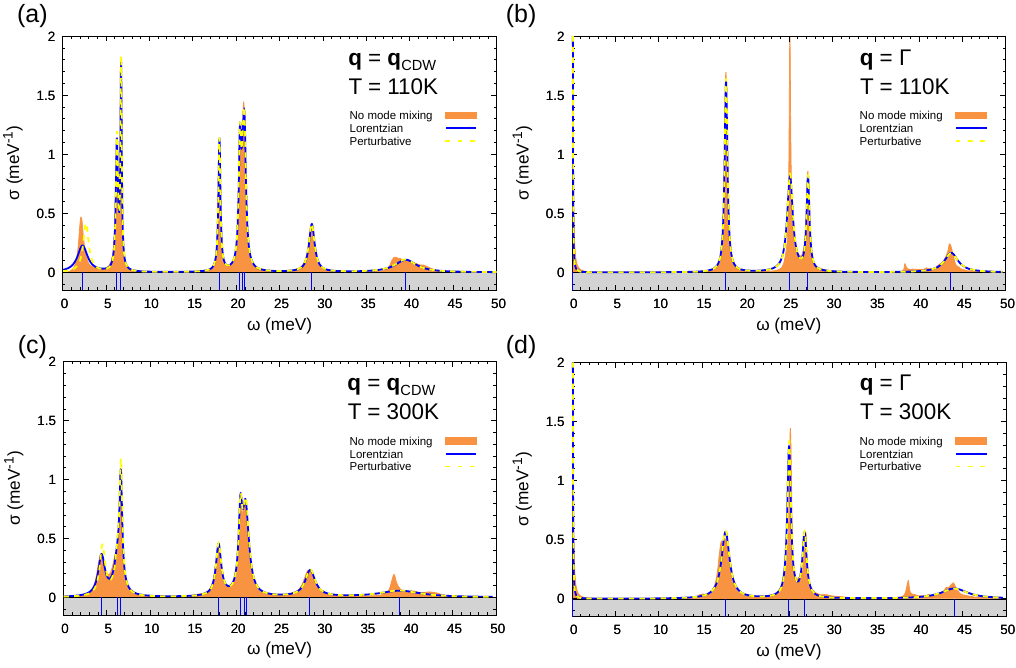}
	\caption{Anharmonic phonon spectral functions at 110 K at (a) \textbf{q}$_{\text{CDW}}$ = (0 1/4 -1/3) r.l.u. and (b) $\Gamma$ points, and at RT at (c) \textbf{q}$_{\text{CDW}}$ and (d) $\Gamma$ points. The orange filled area indicates the result obtained with the ``no mode-mixing'' approximation in Eq. \eqref{nomm}. The blue solid line is the spectrum calculated in the ``one-shot'' Lorentzian approximation in Eqs. \eqref{eq:os_w} and \eqref{eq:os_gamma}. The center of these Lorentzians, specified with vertical blue lines in the grey shaded area below, indicate the anharmonic phonon frequencies in the quasiparticle picture. The dashed yellow line is the spectrum in the  perturbative limit (see Eqs. \eqref{perturbative}).}
	\label{FigAnh}
\end{center}
\end{figure*}

Before moving to experimental results, in this section we present the theoretical analysis of the anharmonic spectral functions $\sigma(\textbf{q}, \omega)$ at ambient pressure at two relevant temperatures including Grimme's semiempirical vdW interactions: the experimental critical temperature (110 K) and RT. First, we check the adequacy of the approximations introduced in Section \ref{dyn} by calculating the anharmonic phonon spectral function at the critical wavevector $\sigma(\textbf{q}_{CDW}, \omega)$ and 110 K. As shown in Fig. \ref{FigAnh} (a), not all modes have a Lorentzian profile. For instance, the highest energy mode has a clear satellite peak, indicating that anharmonicity makes the quasiparticle picture questionable for this mode and, consequently, the Lorentzian approximations do not work well. This result emphasizes that anharmonicity is not only crucial to stabilize the lowest energy mode at \textbf{q}$\mathrm{_{CDW}}$ \cite{Diego2021}, it also impacts high-energy optical modes considerably in their spectral properties. Nevertheless, the spectrum obtained within the Lorentzian ``one-shot'' approximation, plotted with blue solid lines, yields a reasonable result for most modes. In particular, for the lowest energy mode, the one driving the CDW transition, the result is not identical to the non-Lorentzian ``no-mode mixing'' approximation, but in both cases the positions coincide. The spectrum obtained in the  perturbative limit, plotted with dashed yellow lines, is generally in agreement with the ``one-shot'' calculation. However, the frequency attributed to the mode which drives the CDW transition is blue shifted in the latter case. This means the SSCHA self-energy cannot be considered small for the softened mode, and the perturbative approach fails. As shown in Fig. \ref{FigAnh} (c), temperature flattens and broadens all phonon peaks due to enhanced scattering between phonons, hindering their experimental detection. Interestingly, the soft mode is strongly blue shifted with temperature up to ~5 meV at 300 K as expected from the previous static calculations \cite{Diego2021}, while the other two acoustic modes become quasi-degenerate.
 
In Fig. \ref{FigAnh} (b) and (d) the anharmonic phonon spectral function is presented at the $\Gamma$ point for both 110 K and 300 K. For low-energy excitations the phonon quasiparticle picture is rather well defined.
However, the huge phonon satellite ascribed to the highest-energy optical mode confirms this mode is strongly anharmonic throughout the whole BZ despite its high energy and that anharmonicity removes its quasiparticle nature. In fact, the splitting of this peak in two can lead to confusions during experiments. The highest energy mode is infrared (IR) and not Raman active, so that it might be difficult to observe the peak splitting with either of these techniques due to the large reflectivity of metals in the IR.

\begin{figure*}[t]
\begin{center}
\hspace{0.4cm}
\includegraphics[width=1.74\columnwidth]{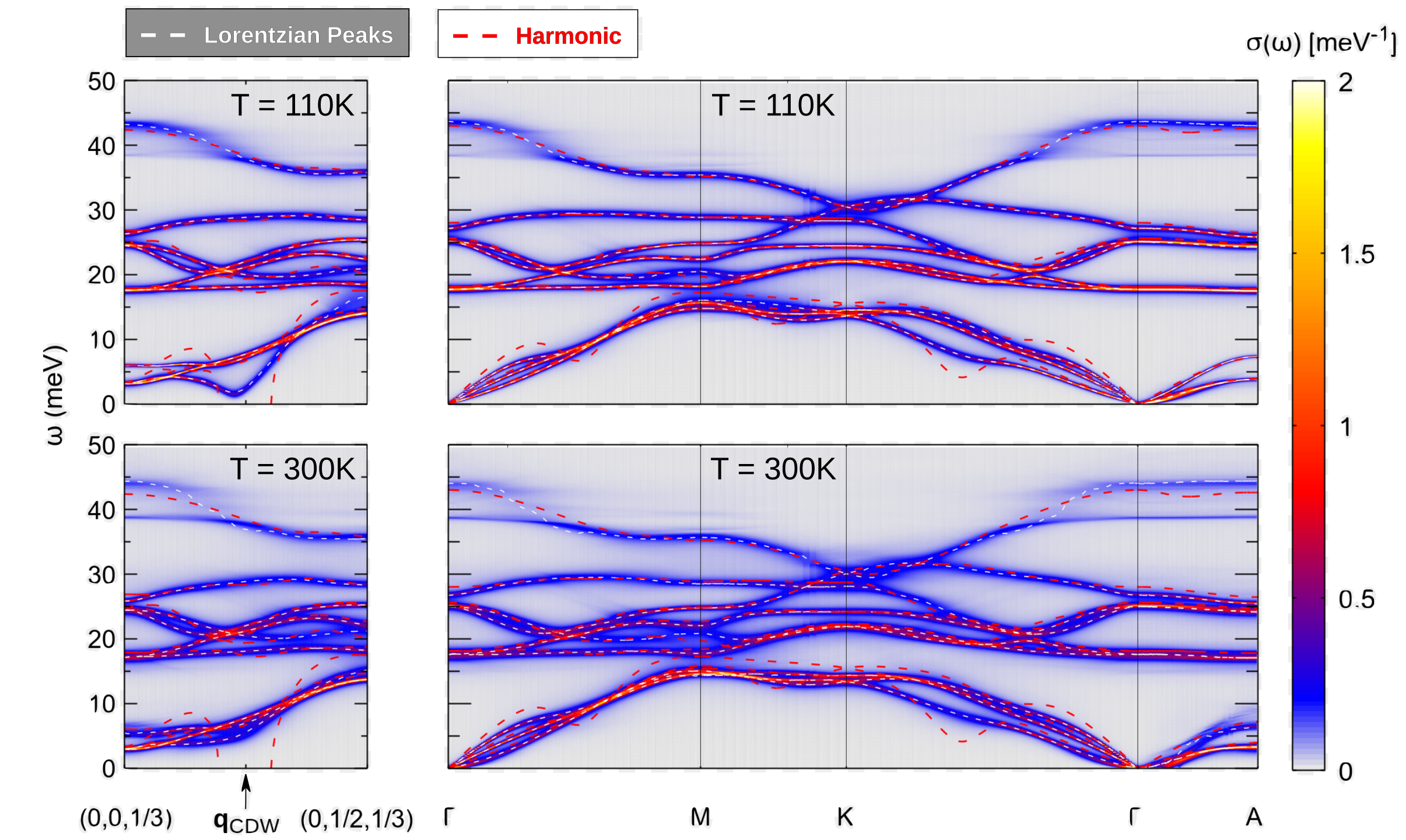}
\caption{Anharmonic phonon spectral functions at 110 K and 300 K along (0 \textit{k} 1/3) r.l.u. and $\Gamma-\mathrm{M-K}-\Gamma-\mathrm{A}$ paths. The dashed white lines are the anharmonic phonon frequencies extracted from the center of the Lorentzians in the ``one-shot'' approximation, while red dashed lines are the harmonic phonons.}
\label{full}
\end{center}
\end{figure*}

In  Fig. \ref{full} the full anharmonic phonon spectrum is plotted both at 110 K and 300 K and is compared to calculations using the harmonic approximation. The figure clearly illustrates that the main anharmonic renormalization of the phonon frequency concentrates around \textbf{q}$\mathrm{_{CDW}}$. The frequency of the CDW driving, $\omega_1$ mode, unstable in the harmonic approximation is strongly temperature dependent as reflected by the spectral function. It is interesting to remark that also halfway between $\Gamma$-M and $\Gamma$-K  an important anharmonic renormalization of the lowest energy acoustic mode is also observed, despite its temperature dependence being weak. The full spectral function shows that the highest energy optical mode shows a clear double peak structure in most of the BZ at low and high temperatures and the highest energy longitudinal acoustic mode shows a satellite peak close to $A$ at 300 K.  

In the rest of the calculations presented here we assume the ``one-shot'' Lorentzian approximation, which allows an estimation of the phonon linewidth that can be compared directly to the experimentally determined one.

\subsubsection{Diffuse scattering}

\begin{figure}[h!] 
\begin{center}
\includegraphics[width=1.0\columnwidth,draft=false]{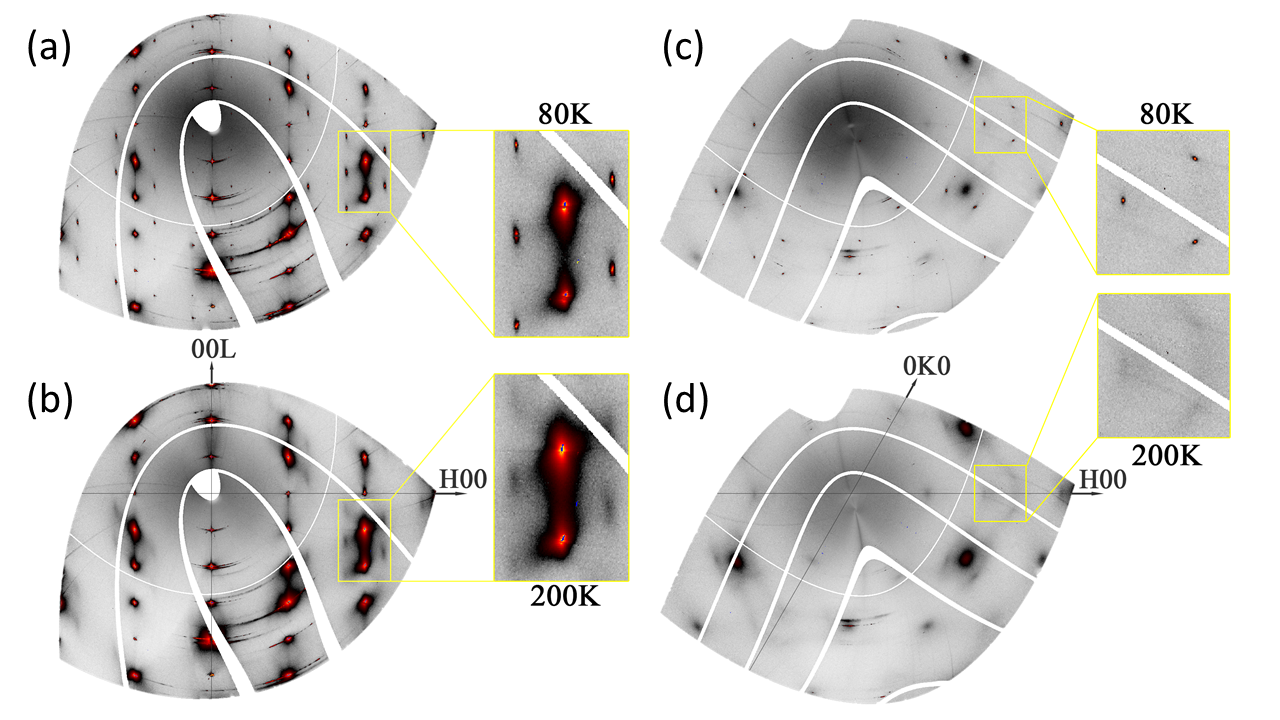}
\caption{Diffuse scattering (DS) maps of VSe$_2$ for the (a) (\textit{h} \textit{0} \textit{l}) and (b) (\textit{h} \textit{k} \textit{0}) planes, showing the CDW reflections at 80 K and diffuse signal at 200 K.}
\label{tds}
\end{center}
\end{figure}

Before discussing the experimental phonon data from VSe$_2$, we describe the diffuse scattering (DS) measurements (Fig. \ref{tds}). Diffuse signal coming from the condensation of a mixture of longitudinal and transverse phonons is already present at temperature above 250 K in the (\textit{h} 0 \textit{l}) plane. The DS develops its maximum intensity at the reciprocal lattice vector $\textbf{G}$$_{2 0 1}$  and follows the \textit{C}$_3$ symmetry of the \textit{P$\overline{3}$m1} space group. This CDW precursor increases its intensity on cooling and condenses into a 3D-CDW at T$\mathrm{_{CDW}}$$\approx$ 110 K, in agreement with the transport and x-ray data of Fig. \ref{Fig1} and \ref{Fig2}.

\subsubsection{Low-energy lattice dynamics: IXS}

\begin{figure}[h!]
\begin{center}
\includegraphics[width=\columnwidth,draft=false]{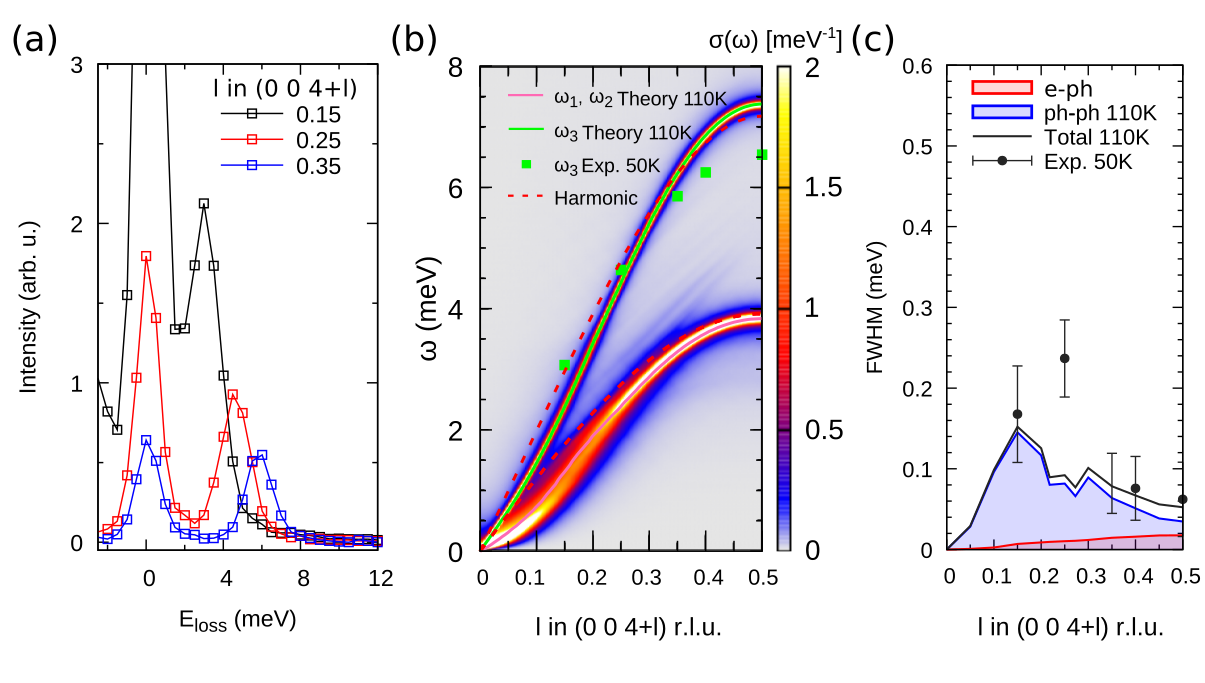}
\caption{(a) Representative momentum dependence of the IXS spectra at 50 K, (b) anharmonic spectral function compared to the experimental peaks and (c) linewidth of the longitudinal (0 0 4+\textit{l}) r.l.u. phonon at 50 K in experiments and 110 K in anharmonic theoretical calculations.
}
\label{004}
\end{center}
\end{figure}

In Fig. \ref{004} we plot the DFT anharmonic spectral function along the $\Gamma$-A direction at 110 K in the high-temperature phase, together with the experimental measurements at 50 K performed along the (0 0 4+\textit{l}) r.l.u. direction in the CDW phase. As the $\Gamma$-A direction lies along the \textit{C}$_{3}$ rotation axis, the only visible acoustic branch $\omega_3$ (higher energy) corresponds to a pure longitudinal mode polarized along the \textit{c} direction of the crystal. The remaining modes $\omega_1$ and $\omega_2$ (which are degenerate in frequency) are transversal and, thus, invisible along this direction. The $\omega_3$ phonon follows a sinusoidal dispersion, consistent with DFT calculations. At 110 K, it has a frequency of 7 meV at the BZ border, lower than the acoustic mode at $\mathrm{M}$, presumably due to the confinement of the acoustic mode in the low dimensional VSe$_2$ layer. This longitudinal c-axis polarized mode is barely temperature dependent and the experimental dispersion measured at 50K, even though it is within the CDW phase, also follows the dispersion obtained from the harmonic calculations too. However, as it can be seen in Fig. \ref{004} (c), the largest contribution to the linewidth of $\omega_3$ comes from anharmonic effects rather than from EPI. Furthermore, as shown in Fig. \ref{full}, we also see that the anharmonic contribution to the total linewidth clearly increases with the temperature. Together, this indicates that anharmonicity has a larger impact on linewidths than frequencies for the phonon modes along the \textit{C}$_{3}$ rotation axis. \\

\begin{figure*}[t]%
\begin{center}
        \includegraphics[width=1.6\columnwidth,draft=false]{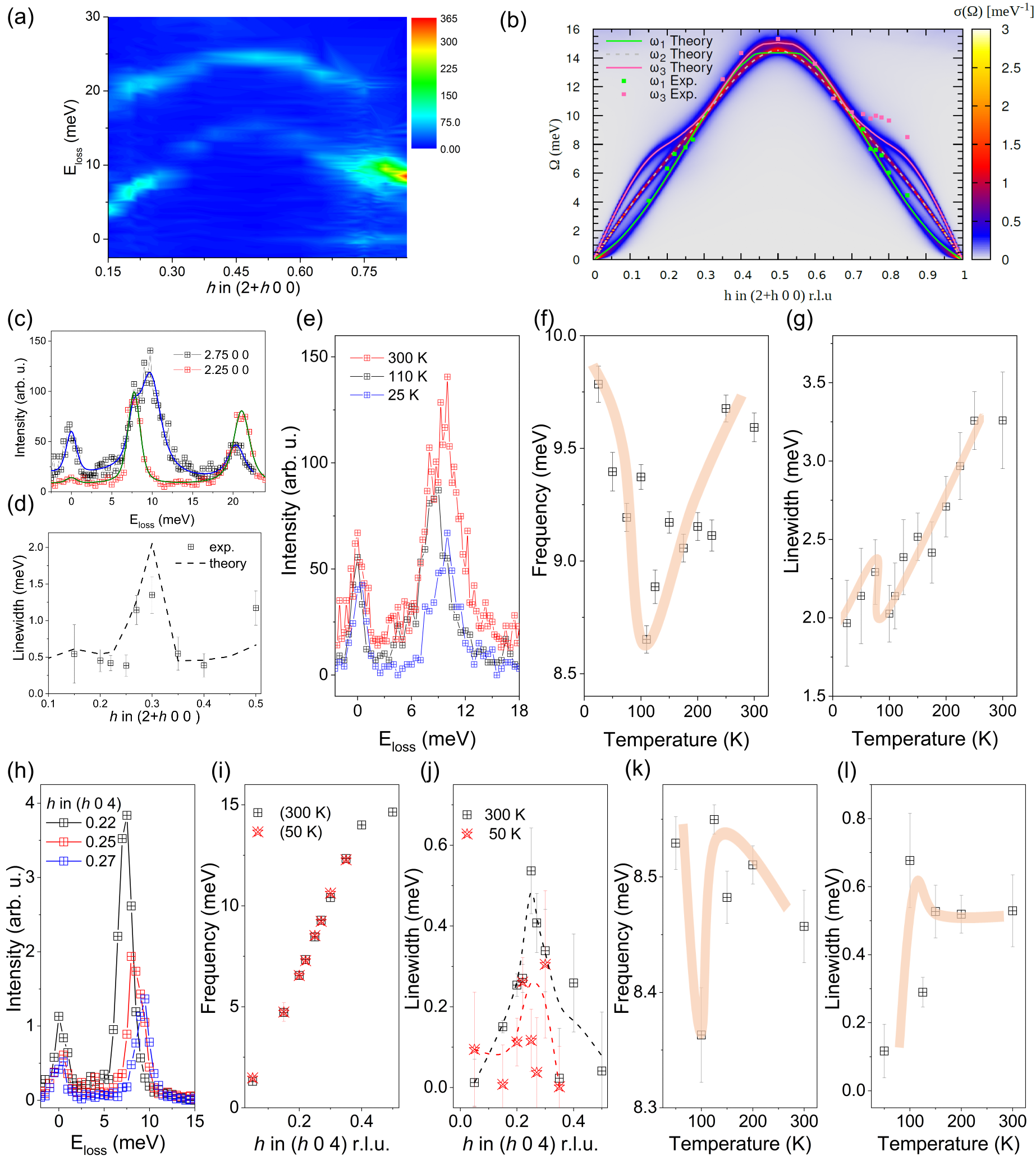}
        \caption{(a) IXS dispersion of the low-energy  phonons at 300 K along the (2+\textit{h} 0 0) direction, for 0.15$<$\textit{h}$<$ 0.85 r.l.u. at 300 K. 
(b) SSCHA spectral function calculated in the Lorentzian ``one-shot'' approximation at 300 K together with the experimental data. (c) Representative IXS spectra for \textit{h}= 0.25 and 0.75 r.l.u., highlighting the overlapping of the 2 longitudinal modes at high \textit{q}. (d) Momentum dependence of the (2+\textit{h} 0 0) linewidth of phonon (full-width-at-half maximum, FWHM) for 0.15$<$\textit{h}$<$0.5 r.l.u. at 300 K both from IXS and anharmonic SSCHA calculations including both phonon-phonon and EPI. (e-g) Temperature dependence of the longitudinal mode at \textit{q}= (2.25 0 0) r.l.u.: raw data (e), frequency (f) and linewidth (FWHM) (g). (h) Raw IXS spectra of the transverse character of the acoustic branch and (i) the corresponding dispersion obtained from the analysis at 300 and 50 K. (j) Momentum dependence of the linewidth of the transversal mode at 300 and 50 K. Broken lines are a guide to eye. (k) Temperature dependence of the frequency and (l) linewidth (FWHM) of the transversal branch $\mathrm{q}$= (0.25 0 4) r.l.u.   Note that the softening in (f) is one order of magnitude larger than in (k).}
        \label{All}
\end{center}
\end{figure*}

\begin{figure}[h!] 
\begin{center}
\includegraphics[width=1.0\columnwidth,draft=false]{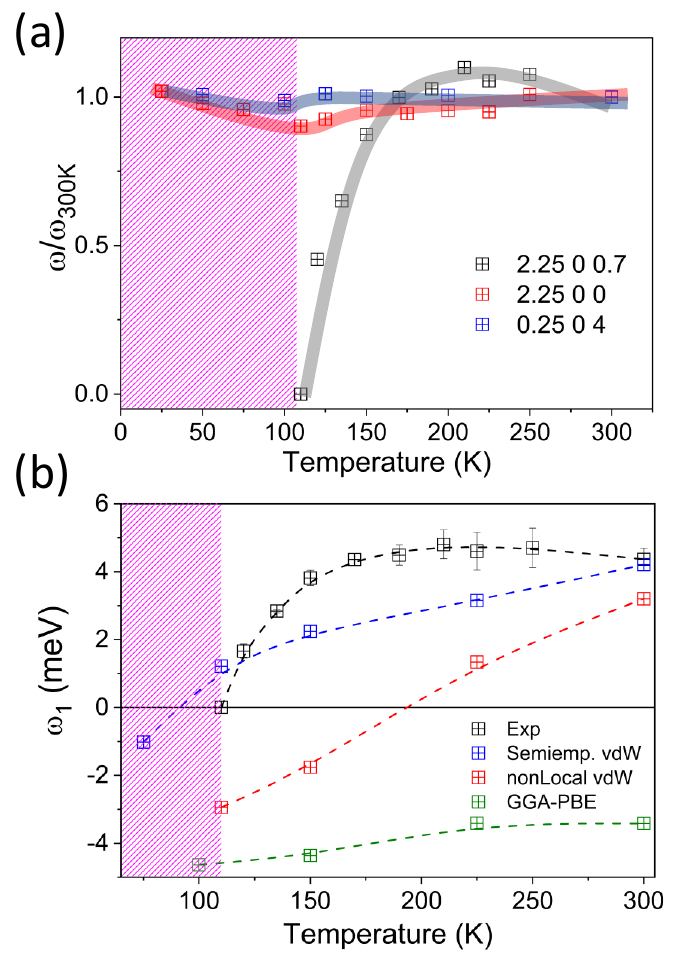}
\caption{(a) Experimental temperature dependence of the $\omega_1$ mode normalized frequency at (2.25 0 0.7) r.l.u., the CDW wavevector, (2.25 0 0) r.l.u. and (2.25 0 4) r.l.u. (b) Experimental temperature of the $\omega_1$ mode energy at (2.25 0 0.7) r.l.u. together with the anharmonic theoretical energies obtained with and without vdW corrections. The shaded area defines the CDW region in both panels.}
\label{vdw}
\end{center}
\end{figure}

We now turn our attention to the phonon branches propagating in the VSe$_2$ plane. Fig. \ref{All} (a) and (b) summarize the experimental data taken along the (2+\textit{h} 0 0) r.l.u. direction for 0.15$<$\textit{h}$<$0.9 ($\Gamma-\mathrm{M}-\Gamma$) at 300 K. The intensity-dispersion of the VSe$_2$ lattice dynamics matches the DFT anharmonic calculation, where two branches are clearly visible below 30 meV. The highest mode corresponds to an optical branch and disperses along the $\Gamma-\mathrm{M}$ direction, with frequencies from 20 to 25 meV. Focusing on the purely acoustic phonon, between 0.15$<$\textit{h}$<$0.5, we observe one single lattice mode, which changes character along the $\Gamma-\mathrm{M}$ direction. Between 0.22$<$\textit{h}$<$0.3, $\omega_1$, with transversal character, it develops more intensity (see Fig. \ref{All} (c), where we plot the (2.25 0 0) r.l.u. scan), while the branch with longitudinal character ($\omega_3$) dominates between 0$<$\textit{h}$<$0.22 and 0.3$<$\textit{h}$<$0.5. Furthermore, the analysis of the linewidth of the acoustic branch along the (2+\textit{h} 0 0) r.l.u. direction (0$<$\textit{h}$<$0.5 r.l.u.) reveals an anomalous broadening around \textit{h}=0.3, which is well captured by our anharmonic calculations. We point out here that the remaining transversal $\omega_2$ mode is silent in IXS. On the other hand, along the 0.5$<$\textit{h}$<$1 direction, both the $\omega_1$ and $\omega_3$ modes can be discerned (see the spectrum at (2.75 0 0) r.l.u. in \ref{All} (c)) and their dispersions are superimposed to the dynamical phonon spectra calculated with the SSCHA in Fig. \ref{All} (b). The $\omega_1$ and $\omega_3$ modes overlap between 0.6$<$\textit{h}$<$0.7 (Fig. \ref{All} (c)), thus, were fitted to a single Lorentzian convoluted with the experimental resolution. Furthermore as is directly visualized in the raw spectra, the temperature dependence of the frequency of $\omega_3$ undergoes a partial softening ($\sim$ 1.3 meV), which is accompanied by a small broadening of the total linewidth ($\omega_1$ + $\omega_3$) at T$\mathrm{_{CDW}}$ (Fig. \ref{All} (f) and (g)), reflecting the response of the acoustic branch to the opening of the CDW gap.  In Fig. \ref{All} (h), we plot a representative set of the single phonon dispersion along the (\textit{h} 0 4) r.l.u. direction (0$<$\textit{h}$<$0.5), which probes only the transversal component of the acoustic mode. It disperses linearly up to 15 meV and shows again a small softening and broadening at \textit{h}=0.25 and 110 K (Fig. \ref{All} (j)) followed by a phonon hardening and narrowing down to 50 K. The temperature dependence of the transversal, longitudinal and the soft branches are summarized in Fig. \ref{vdw} (a). We can clearly observe that, while the phonon at (2.25 0 0.7) r.l.u. fully collapses at T$\mathrm{_{CDW}}$ \cite{Diego2021}, the transversal and longitudinal branches soften by only 12\% and 3\%, respectively. This is in agreement with the ARPES data shown in Fig. \ref{Fig3} (c), which shows a depletion of the DOS at an in-plane propagation vector consistent with the partial phonon softening as well as along the whole \textit{k}$_z$ direction. 

\onecolumngrid
\twocolumngrid

Figure \ref{vdw} (b) compares the experimental data with the SSCHA calculations using different exchange-correlation functionals for the vdW interactions.
We performed SSCHA minimizations \textit{ab initio} with and without considering vdW corrections at all the temperatures at which the IXS experiments were carried out. We center our analysis on the characterization of the CDW transition through the study of the temperature dependence of the softest acoustic mode at \textbf{q}$\mathrm{_{CDW}}$. As explained in Section \ref{dyn}, from a thermodynamic point of view, the structural instability under study can be correctly characterized through the phonons obtained by diagonalizing the free energy Hessian. As already discussed in Ref. \cite{Diego2021}, anharmonicity stabilizes the 1\textit{T} structure of VSe$_2$ at high temperatures only if vdW forces are taken into account in the calculation of the energies and forces needed to carry out the SSCHA variational minimization.  vdW corrections must need included in order to characterize properly the transition that is related to the out-of-plane nature of the CDW, in which the interlayer distance is modulated. Definitely, anharmonicity and weak vdW interactions are responsible for the melting of the CDW.

The calculated CDW critical temperature strongly depends on how vdW interactions are included. The best agreement with the experimental result is obtained with the DFT-D2 method of Grimme \cite{doi:10.1002/jcc.20495}, plotted with blue squares in Fig. \ref{vdw}, which simply adds a semiempirical dispersion correction on top of a converged Kohn-Sham energy. This theoretical result predicts that the soft mode frequency vanishes between 75 and 110 K, in  good agreement with the experimental value of T$\mathrm{_{CDW}}$ = 110 K \cite{Diego2021}. The, a priori, more sophisticated vdW-DF functional proposed by Dion et al. \cite{PhysRevLett.92.246401}, which is truly non local, is plotted with red squares and succeeds in melting the CDW, but overestimates the critical temperature by approximately 80 K. The lower critical temperatures obtained with Grimme's semiempirical vdW interactions might be due to the fact that it provides larger forces than the non-local vdW functional. It is interesting to remark, however, that the non-local functional vdW functional reproduces the CDW physics much better than the semiempirical approach in the VSe$_2$ monolayer \cite{Fumega2023}, which is unable to predict the existence of the CDW transitions. The reason may be that Grimme’s semiempirical approach is rather elemental and probably is introducing some arguably spurious intralayer contributions in supercell forces and energies that is slightly overestimating the contribution of vdW corrections. 

\subsection{Anisotropic EPI}

The origin and stabilization mechanism of the CDW transition is still under debate, especially in systems with dimensions higher than 1D or quasi-1D, where Peierls' initial analysis of the Fermi surface nesting scenario probably does not hold. In fact, IXS and theoretical calculations already highlighted the importance of a highly momentum dependent EPI in the formation of the CDW in similar compounds of the TMD family, such as 2\textit{H}-NbSe$_2$ \cite{Web11Nb,PhysRevB.80.241108} and 1\textit{T}-TiSe$_2$ \cite{Web11Ti,Calandra2011}. Regarding 1\textit{T}-VSe$_2$, the critical role of the momentum dependence of the EPI was already pointed out in \cite{Henke2020} by means of quantitative models. More recently, IXS experiments and DFT-based SSCHA calculations have agreed in endorsing the electron-phonon coupling as the driving mechanism of the CDW transition in 1\textit{T}-VSe$_2$, in spite of the presence of nesting \cite{Diego2021}. 

Despite reaching clear conclusions, these calculations were only done for the constant height path in the reciprocal space that passes through \textbf{q}$\mathrm{_{CDW}}$. Considering that experiments trying to elucidate the CDW origin base their conclusions on the width of the BZ range in which the phonon branch is damped, we consider useful to provide a deeper analysis by extending both the experimental and theoretical analysis to the perpendicular directions (\textit{h k l}) r.l.u. around \textbf{q}$\mathrm{_{CDW}}$.

\begin{figure} 
\begin{center}
\includegraphics[width=\columnwidth,draft=false]{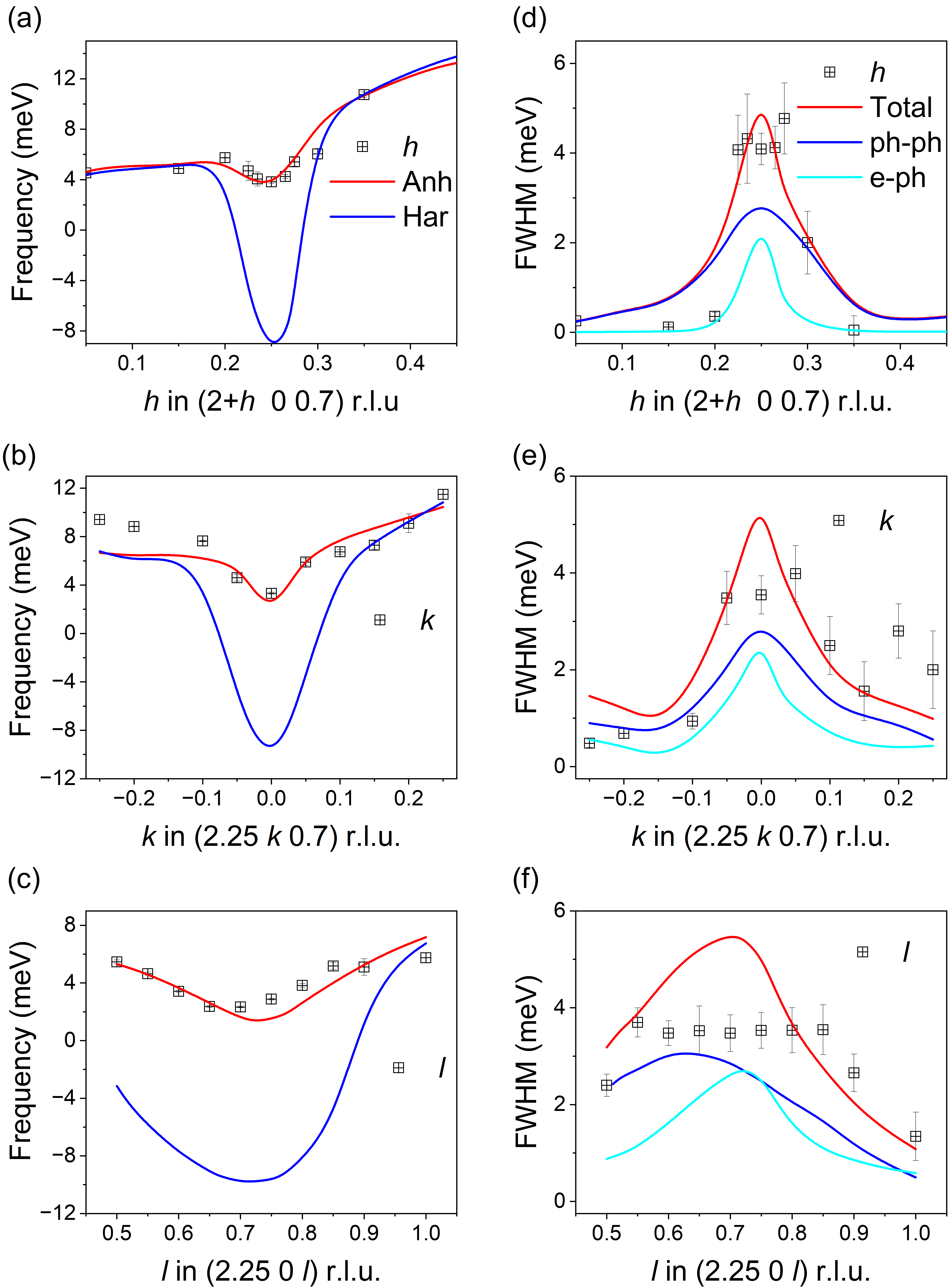}
\caption{(a-c) Experimental (points) and calculated (solid lines) dispersion of the soft mode at 150K around \textbf{q}$\mathrm{_{CDW}}$, but probing perpendicular directions (\textit{h k l}) r.l.u. (d-f) Experimental (points) and theoretical (solid lines) linewidth as a function of momentum. The calculated linewidth (red line) includes the contribution of the EPI (light blue line) and anharmonicity (dark blue line).}
\label{Aniso}
\end{center}
\end{figure}

In Fig. \ref{Aniso}, we compare the frequencies of the soft phonon branch together with their total linewidth, which includes both the electron-phonon and anharmonic contributions, at 150 K, sweeping the CDW critical wavevector along \textit{h}, \textit{k} and \textit{l}. Figures \ref{Aniso} (a-c) compare the momentum dependence of the phonon softening along the three directions with the (an-)harmonic calculations. The theoretical calculations nicely match the experimental data when considering  anharmonicity within the SSCHA. Similarly, the experimental linewidth, which shows an anomalous broadening at the critical wavevector, follows the total linewidth obtained theoretically. Considering this effect in finer detail, we find that the region of momenta that undergoes a phonon renormalization is anisotropic in momentum space, both experimentally and theoretically. The phonon softening (broadening) extends to 0.075 r.l.u. (0.02 \r{A}$^{-1}$) along \textit{h}, 0.016 r.l.u. (0.045 \r{A}$^{-1}$) along \textit{k}, and 0.35 r.l.u. (0.06 \r{A}$^{-1}$) along \textit{l}. This momentum dependence is already present at the harmonic level, which reflects the anisotropy of the electron-phonon matrix-elements in momentum space, and nesting in a lower extent. 

Considering the Fermi surface nesting scenario as the driving force of the CDW formation, naively one would expect an isotropic softening (broadening) as a consequence of the localization of the phonon fluctuations in momentum space. This is in clear contrast with the data reported in Fig. \ref{Aniso}.
The anisotropic behaviour we report evinces the fragility of proposing the nesting scenario as the origin of the CDW based only on the width
of Brillouin zone range in which the phonon branch is damped. Precisely, this kind of arguments may be conditioned by the direction in which the scan is performed, at least in the case of 1\textit{T}-VSe$_2$. Still, the fact that the anisotropy is accentuated in the out-of-plane \textit{l} direction, suggests that it may be related to the quasi-2D nature of the compound and, thus, it is reasonable to think that the EPI and anharmonic effects are also anisotropic in other compounds of the TMD family and in other strongly correlated electron systems. Regardless of any potential generality, our results allow us to indicate that a pure nesting scenario for 1\textit{T}-VSe$_2$ cannot solely act the driving force, and clearly support previous claims that point to the electron-phonon coupling as the trigger for the CDW formation.

\subsection{Pressure dependence of the soft mode at RT}

Having comprehensively described the electronic structure and the temperature dependence of the low-energy lattice dynamics of VSe$_2$, we finish our experimental survey by studying the evolution of the soft mode with pressure at RT. Systematic studies of the pressure dependence of the CDW in 2D and correlated materials reported a dramatic weakening of the charge modulations under low-to-moderate pressure \cite{Leroux2015,Souliou2014,Li2022,Chen2021,Wang2021}. In particular, the suppression of the CDW in 2\textit{H}-NbSe$_2$ is directly related to the strong anharmonic character of the lattice potential, which stabilizes the high temperature phase under pressure \cite{Leroux2015}. 
Recent transport experiments reported an intriguing enhancement of the CDW order in VSe$_2$ up to almost RT at $\sim$13 GPa, in clear contrast to the typical behaviour observed in CDW systems. At this pressure 1$T$-VSe$_2$ undergoes a first order phase transition to a new $C2/m$ phase,  which destroys the charge order and allows superconductivity to emerge \cite{Sahoo2020}. Furthermore, high-pressure Raman \cite{Feng2020}, x-ray diffraction, and spectroscopic experiments \cite{Guo2021} confirmed the transition to the $C2/m$ phase proceeds at RT, similar to the 1\textit{T}´-structures of ditellurides (Nb,Ta)Te$_2$ \cite{Lin2022}. 

\begin{figure} 
\begin{center}
\includegraphics[width=1.0\columnwidth,draft=false]{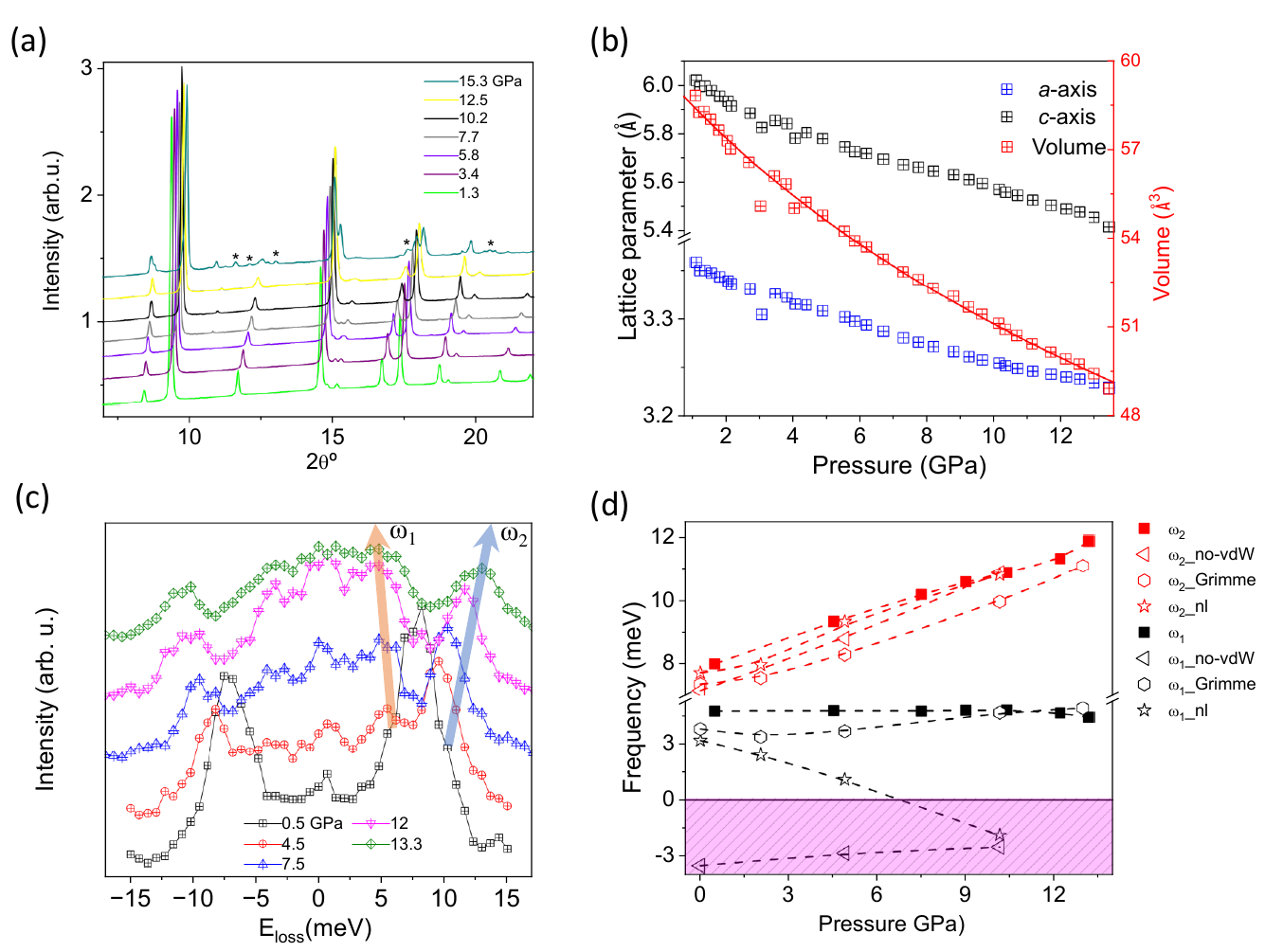}
\caption{(a) Representative x-ray diffraction scans as a function of pressure at 300 K. Peaks marked with an asterisk denote the new phase. (b) Pressure dependence of the lattice parameters \textit{a} and \textit{c}. (c) Experimental pressure dependence of $\omega_1$ and $\omega_2$ at $\mathbf{q}_{\mathrm{CDW}}$ and (d) its comparison with SSCHA results from the free energy Hessian including different vdW functionals.}
\label{pressure_xrd}
\end{center}
\end{figure}

Figure \ref{pressure_xrd} summarizes the pressure experiments carried out at RT. Up to 13 GPa, Fig. \ref{pressure_xrd}(a), the x-ray diffraction data shows a gradual displacement of the Bragg peaks towards larger detector angles, signalling the decrease of the lattice parameters under pressure. New Bragg peaks corresponding to a different phase, labelled with asterisks, appear above 13 GPa. The diffraction patterns below 13 GPa (\textit{P$\overline{3}$m1} space group) were fitted with the GSAS software and returned the evolution of the lattice parameters displayed in Fig. \ref{pressure_xrd}(b). Fitting the pressure dependence of the volume to the second order Birch-Murnaghan equation of state (EoS), 
\begin{equation}
\begin{aligned}
    P\left(V\right) = & \frac{3B_0}{2}\left[\left(\frac{V_0}{V}\right)^{7/3}-\left(\frac{V_0}{V}\right)^{5/3}\right]\\
    & \left\{1+\frac{3}{4}\left(B_0'-4\right)\left[\left(\frac{V_0}{V}\right)^{2/3}-1\right]\right\},
\end{aligned}
\end{equation}
where $\mathrm{P}$ is the pressure, $\mathrm{V}_0$ is the volume at zero pressure, $\mathrm{B}_0$ is the Bulk modulus, and $\mathrm{B}_0$´ is the derivative of the bulk modulus with respect to pressure,
returns a bulk modulus of $\mathrm{B}_0 =46.5$ GPa, slightly larger than the value reported in the literature \cite{Song2021}.    

Due to a favored scattering geometry, high-pressure IXS spectra of VSe$_2$ were recorded in the BZ adjacent to the $\textbf{G}_{0 3 1}$ reciprocal lattice vector (0 2.25 0.7) r.l.u., Fig. \ref{pressure_xrd}(c)), equivalent to the $\textbf{G}_{2 0 1}$ (see the diffuse scattering in Fig. \ref{tds}), scanning in reciprocal space along \textit{k} as the pressure was increased.
The results of these fits are given in Fig. \ref{pressure_xrd} (d). Focusing on the $\omega_2$ branch, its frequency increases linearly under pressure ($\sim$ 0.3 meV/GPa) up to 13 GPa, as expected as the lattice becomes stiffer under pressure. Further, the linewidth decreases up to 10 GPa, after which the mode broadens on approaching the high pressure phase transition (not shown). In contrast, the frequency of the soft $\omega_1$ mode remains pressure independent up to 10 GPa, without collapsing on approaching the high pressure phase, but softens $\sim$10$\%$ between 10 and 13 GPa. The phonon linewidth of this mode smoothly increases with pressure. The distinct behaviour of the $\omega_1$ phonon with temperature and pressure demonstrates the different nature of the phase transitions. While the CDW transition at 110 K agrees with a mean field behaviour with 0.5 critical exponent, the RT-high pressure transition is better described by a first order phase transition. 

SSCHA anharmonic calculations have been performed at RT for pressure values in the experimental range, making use of the experimental lattice parameters in Fig. \ref{pressure_xrd} (b). The application of pressure leads to a reduction of the interlayer distance $c$, which in turn lowers out-of-plane vdW interactions. Consequently the choice of exchange-correlation functional has to be performed carefully. The theoretical phonon frequencies calculated with different exchange-correlation functionals are presented in Fig. \ref{pressure_xrd} (d). GGA-PBE wrongly predicts that, at 300 K, the system stays in the CDW phase for all pressures. The non-local vdW exchange-correlation functional captures the CDW transition under pressure as the $\omega_1$ phonon at $\mathbf{q}_{\mathrm{CDW}}$ softens with increasing pressure, indicating that this functional correctly describes the increase of the CDW critical temperature with pressure. However, it underestimates the transition pressure to the new phase, in agreement with the fact that it overestimates at 0 GPa the CDW critical temperature. The semiempirical approach yields a phonon frequency $\omega_1$ that increases with pressure, the opposite to the expected behaviour and the experimentally observed trend, even if the absolute values obtained with this approach yield the closest frequencies to the experimental ones. 
All functionals yield accurate frequencies for the  $\omega_2$ phonon branch. The behaviour of the frequencies with respect the choice of functionals resembles the situation in the monolayer limit \cite{Fumega2023}, in which the inclusion of the non-local vdW interactions in the exchange-correlation functional was mandatory to obtain a good description of the transition to the CDW phase. 

\section{Discussion and conclusions}
Before summarizing the conclusions of our work, we will briefly restate the major results. The resonant x-ray diffraction does not find any possible indication of either chiral or orbital order, as reported for the 1\textit{T}-TiSe$_2$ system \cite{Wezel2011,Castellan2013,Ishioka2010,Ueda2021,Xu2020,Peng2022}. We see that a CDW with propagation vector (0.25 0 -0.3) r.l.u. in 1\textit{T}-VSe$_2$ opens a pseudogap at T$>$T$_C$, with further depletion of DOS below T$_C$. Autocorrelation analysis of the Fermi surface maps finds indications of an enhanced charge susceptibility with propagation vector similar to the experimentally observed \textbf{q}$_\mathrm{CDW}$, but spread across all k$_z$ values. It turns out that the maximum enhanced charge susceptibility coincides with the phonon collapse at  T$\mathrm{_{CDW}}$. However, from a nesting scenario point of view \cite{Johannes2006}, one would expect (\blue{1}) a singularity in the charge correlation map, which isotropically localizes the phonon fluctuations in momentum space, (\blue{2}) a collapse of the low-energy phonon in a very narrow region of momenta \cite{Hoesch2011} and (\blue{3}) an isotropic EPI, contrary to the experimental observations. Furthermore, the acoustic longitudinal and transversal modes with \textit{l}=0 and 0.5 also undergo phonon renormalizations without collapsing, hence correlating the phonon anomalies with the charge susceptibility. In fact, the anisotropic linewidth shown in Fig. \ref{Aniso} is derived from highly anisotropic electron-phonon coupling matrix elements, which are related to the symmetry of the phonon polarization and the anisotropic electronic band structure.

\begin{figure}[t] 
\begin{center}
\includegraphics[width=\columnwidth,draft=false]{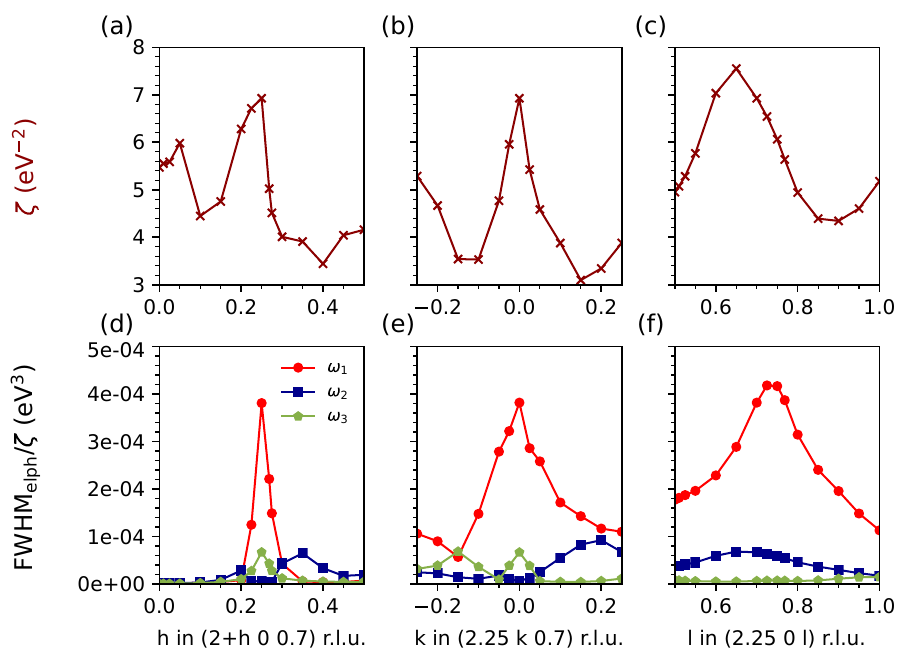}
\caption{(a-c) Nesting function around \textbf{q}$\mathrm{_{CDW}}$, probing perpendicular directions (\textit{h k l}) r.l.u. (d-f) Ratio between the full width at half maximum given by the electron–phonon interaction and the nesting function at the same wavevectors.}
\label{conclusions}
\end{center}
\end{figure}

To gain a deeper understanding of the correlation between the electronic structure and lattice dynamics of VSe$_2$,  we compute the nesting function $\zeta$(\textbf{q)}:
\begin{equation}
    \zeta(\textbf{q})=\frac{1}{N}\sum_{nn'}\sum_{\textbf{k}}^{1BZ} \delta(\epsilon_{n'\textbf{k}+\textbf{q}})\delta(\epsilon_{n\textbf{k}}) \;.\label{eq:nesting}
\end{equation}
The nesting function probes the Fermi surface by peaking at nesting \textbf{q} wavevectors, revealing if the CDW emerges from a purely electronic instability. In Fig. \ref{conclusions} (a-c) we plot the nesting function around the perpendicular directions (\textit{h k l}) r.l.u. of \textbf{q}$\mathrm{_{CDW}}$. The nesting function peaks just at the CDW vector for \textit{h} and \textit{k} in-plane scans, and nearby for the out-of-plane \textit{l} scan, so that \textbf{q}$\mathrm{_{CDW}}$ coincides with a nested region of the Fermi surface. 

In order to compare nesting and EPI scenarios, we plot the ratio between the electron-phonon linewidth and the nesting function in Fig. \ref{conclusions} (d-f), which allows us to assess directly the momentum and mode dependence of the electron–phonon coupling matrix elements following Eq. \eqref{eq:elph}.  
The ratio shows that the EPI matrix elements still depend  more strongly on momentum than the nesting function, and in fact, they have a very similar shape to the electron-phonon linewidth plotted in grey in Fig. \ref{Aniso}. Remarkably, the scan alongside the out-of-plane direction shows that while the nesting function favours a commensurate $4 \times 4 \times 3$ CDW reconstruction, it is the EPI mechanism that leads to the incommensurate CDW reconstruction reported in our experiments with $l\simeq0.7$ r.l.u. Therefore, despite the presence of Fermi surface nesting, these calculations also suggest that it is the EPI that is mostly responsible for the formation of the CDW in VSe$_2$.

In conclusion, the thorough experimental and theoretical analysis further supports the argument that the EPI that is the main driving force of the CDW transition in 1$T$-VSe$_2$. This is in agreement with the general consensus that is being built in TMDs based on inelastic scattering experiments \cite{Web11Nb,Web11Ti,Diego2021}, detailed DFT calculations \cite{Johannes2006,Johannes2008,Henke2020} and recent ARPES data \cite{Weber2018} where the weak FS nesting is rather a factor enhancing the electron-phonon matrix elements than a driver of the CDW formation in these compounds. 

\section{Acknowledgments}
J.D. thanks the Department of Education of the Basque Government for a predoctoral fellowship (Grant No. PRE-2020-1-0220). V.P. and S. B-C acknowledge financial support from  the MINECO of Spain through the grants PID2021-122609NB-C22 and PID2021-122609NB-C21, respectively. S.B-C is also supported by MCIN and by the European Union Next Generation EU/PRTR-C17.I1, as well as by IKUR Strategy under the collaboration agreement between Ikerbasque Foundation and DIPC on behalf of the Department of Education of the Basque Government. I.E. acknowledges funding from the Department of Education, Universities and Research of the Eusko Jaurlaritza, and the University of the Basque Country UPV/EHU (Grant No. IT1527-22). C.P. acknowledges the financial support from MINECO project PID2021-125927NB-C21. A.O.F. thanks the Academy of Finland Project No. 349696. M. C. is co-funded by the European Research Council (ERC, DELIGHT, 101052708). Views and opinions expressed are however those of the author(s) only and do not necessarily reflect those of the European Union or the European Research Council. Neither the European Union nor the granting authority can be held responsible for them.
We thank the European Synchrotron Radiation Facility (ESRF) for provision of synchrotron radiation facilities under proposal numbers HC-4941 and HC-4873. This research used resources of the Advanced Photon Source, a U.S. Department of Energy (DOE) Office of Science user facility operated for the DOE Office of Science by Argonne National Laboratory under Contract No. DE-AC02-06CH11357. LOREA beamline at ALBA Synchrotron Light Source was cofunded by the European Regional Development Fund (ERDF) within the Framework of the Smart Growth Operative Programme 2014-2020. We also thank Jordi Prat for the invaluable technical support during ARPES experiments at LOREA.

\bibliography{VSe2}

\end{document}